\documentclass{aa}
\usepackage{txfonts,graphicx,natbib,epsfig,sidecap}
\usepackage{rotating}
\usepackage{fancybox,amssymb,graphics,subfigure}
\begin{document}
\title{The XMM large scale structure survey: optical vs. X-ray classifications of active galactic nuclei
and the unified scheme}
\author{O. Garcet\inst{1}, P. Gandhi\inst{2}, E. Gosset\inst{1},
P. G. Sprimont\inst{1}, J. Surdej\inst{1}, V. Borkowski\inst{1},
M. Tajer\inst{3,4}, F. Pacaud\inst{5,6}, M. Pierre\inst{5}, L.
Chiappetti\inst{7}, D. Maccagni\inst{7}, M. J. Page\inst{8}, F. J.
Carrera\inst{9}, J.A. Tedds\inst{10}, S. Mateos\inst{10}, M.
Krumpe\inst{11},  T. Contini\inst{12}, A. Corral\inst{9}, J.
Ebrero\inst{9}, I. Gavignaud\inst{11},  A. Schwope\inst{11}, O. Le
F\`{e}vre\inst{13}, M. Polletta\inst{14}, S. Rosen\inst{8}, C.
Lonsdale\inst{14} , M. Watson\inst{10}, W. Borczyk\inst{15}, P.
Vaisanen\inst{16}}

\institute{ Institut d'Astrophysique et de G\'{e}ophysique,
Universit\'{e} de Li\`{e}ge, Belgium \and RIKEN Cosmic Radiation
Lab, 2-1 Hirosawa, Wakoshi, Saitama, 351-0198, Japan \and
INAF-Osservatorio di Brera, via Brera 28, I-20121 Milano, Italy
\and Universit\`{a} degli Studi di Milano - Bicocca, Dipartimento
di Fisica, Piazza della Scienza 3, 20126 Milano, Italy \and
CEA/DSM/DAPNIA Service d'astrophysique, Saclay, 91191 Gif sur
Yvette, France \and Argelander-Institut für Astronomie, Universität Bonn,
Auf dem Hügel 71, 53121 Bonn, Germany \and INAF-IASF Milano, via Bassini 15, I-20133
Milano, Italy \and Mullard Space Science Laboratory, University
College London, Holmbury St Mary, Dorking, Surrey, RH5 6NT, UK
\and Instituto de F\'{i}sica de Cantabria (CSIC-UC), 39005
Santander, Spain \and Department of Physics and Astronomy,
University of Leicester, LE1 7RH, UK \and Astrophysikalisches
Institut Potsdam, An der Sternwarte 16, D-14482 Potsdam, Germany
\and Laboratoire d'Astrophysique de Toulouse et Tarbes (UMR 5572),
Observatoire Midi-Pyr\'{e}n\'{e}es, 14 Avenue E. Belin, F-31400
Toulouse, France  \and Laboratoire d'Astrophysique de Marseille,
UMR 6110 CNRS-Universit\'{e} de Provence, BP8, 13376 Marseille
Cedex 12, France \and Center for Astrophysics \& Space Sciences,
University of California, San Diego, La Jolla, CA 92093-0424, USA
\and Astronomical Observatory of A. Mickiewicz University,
Sloneczna 36, Pozna\'{n} 60-286, Poland \and South African
Astronomical Observatory, Observatory 7935, South Africa}


\titlerunning{The XMM large scale structure survey : Optical vs. X-ray classifications of AGN}
\authorrunning{Garcet et al.}

\date{Received / Accepted}

\abstract{}{Our goal is to characterize AGN populations by
comparing their X-ray and optical classifications within the
framework of the standard orientation-based unified scheme.}{We
present a sample of 99 spectroscopically identified ($R$$\leq 22$
mag) X-ray selected point sources in the XMM-LSS survey which are
significantly detected ($\geq3$ $\sigma$) in the [2-10] keV band
with fluxes between 8$\times$10$^{-15}$ and 8$\times$10$^{-14}$
erg s$^{-1}$ cm$^{-2}$, and which have more than 80 counts. We
have compared their X-ray and optical classifications. To this
end, we performed an X-ray spectral analysis for all of these 99
X-ray sources in order to assess whether they are intrinsically
absorbed or not. The X-ray classification is based on the measured
intrinsic column density. The optical classification is based on
the measured FWHM of the permitted emission lines, the absence of
broad lines being due to obscuration within the framework of the
standard AGN unified scheme.}{Introducing the fourfold point
correlation coefficient $r$, we find a mild correlation between
the X-ray and the optical classifications ($r$=0.28), as up to 32
X-ray sources out of 99 have differing X-ray and optical
classifications: on one hand, 10\% of the type 1 sources (7/32)
present broad emission lines in their optical spectra and strong
absorption ($N$$_{\mathrm{H}}$$^{\mathrm{int}}$$\geq$ 10$^{22}$
cm$^{-2}$) in the X-rays. These objects are highly luminous AGN
lying at high redshift and thus dilution effects by the host
galaxy light are totally ruled out, their discrepant nature being
an intrinsic property instead. Their X-ray luminosities and
redshifts distributions are consistent with those of the
unabsorbed X-ray sources with broad emission lines
($L$$_{2-10}$$\sim$4$\times$10$^{44}$ erg s$^{-1}$; z$\sim$1.9).
On the other hand, 25/32 are moderate luminosity
($L$$_{2-10}$$\leq$ 5$\times$10$^{43}$ erg s$^{-1}$) AGN, which
are both unabsorbed in the X-rays and only present narrow emission
lines in their optical spectra. Based on their line ratios in the
optical, the majority of them have an optical spectrum which is
more representative of the host galaxy rather than of a reddened
AGN. We finally infer that dilution of the AGN by the host galaxy
seems to account for their nature. 5/25 have been defined as
Seyfert 2 based on their optical spectra. In conclusion, most of
these 32 discrepant cases can be accounted for by the standard AGN
unified scheme, as its predictions are not met for only 12\% of
the 99 X-ray sources.}{}

\keywords{X-ray: galaxies - galaxies: active - surveys - galaxies:
quasars}

\maketitle

\section{Introduction}

More than 40 years after its discovery, the X-ray background (XRB)
is partially resolved into discrete sources and its main component
is widely interpreted as being mostly made of AGN (Setti \&
Woltjer 1989, Giacconi et al. 2002, Alexander et al. 2003).
Recently, around 80\% of the XRB has been resolved in the [2-10]
keV energy range by deep \textit{Chandra} and \textit{XMM-Newton}
observations (e.g. Worsley et al. 2005; Hickox \& Markevitch 2006;
Carrera et al. 2007). XRB synthesis models with a combination of
absorbed and unabsorbed AGN and founded on the AGN
orientation-based unified scheme (Antonucci 1993) in which each
AGN contains an obscuring torus, have been able to reproduce the
overall broadband spectral shape of the observed XRB (e.g Comastri
et al. 1995, 2001; Gandhi \& Fabian 2003; Treister \& Urry 2005;
Gilli et al. 2007). The standard orientation-based unification
scheme rather suggests that absorbing material (the putative
torus) is present or not along the line of sight depending on the
relative inclination of the torus. However, a certain number of
observations do not agree with the predictions of these synthesis
models (in particular with the unified scheme). There is a
significant number of AGN for which the expected optical and X-ray
characteristics are not the same, and are thus violating the
unified scheme. On one hand, there are AGN showing strong
absorption in the X-rays while their optical and ultraviolet
emission shows mild extinction (Fiore et al. 1999; Maiolino et al.
2001a; Page et al. 2001; Gallagher et al. 2006). On the other
hand, there are also AGN which are only showing narrow emission
lines in their optical spectra (which would happen e.g. if the
broad line region was hidden), while only mild (or even absent)
absorption is observed in their X-ray spectra (e.g. Pappa et al.
2001; Panessa \& Bassani 2002; Barcons et al. 2003). Finally, very
recently, Punsly (2006) has used a sample including both obscured
quasars and broad absorption line quasars (BAL QSOs) to compare
their hydrogen column density derived from X-ray observations.
They showed that, surprisingly, the BAL QSOs have column densities
that are significantly larger than those of the obscured QSOs,
which is at odds with the AGN unified scheme. In this paper, we
characterize AGN populations over the XMM-LSS area by comparing
their X-ray and optical classifications.

The XMM Large Scale Structure Survey (XMM-LSS) presently consists
of 19 guaranteed-time (G) and 32 guest-observer time (B)
overlapping pointings covering a total area of 6 deg$^{2}$. The
nominal exposure times were 20 ks and 10 ks for the G and B
pointings, respectively. We refer to Pierre et al. (2004) and
Pierre et al. (2007) for details on the X-ray observations.
Details of the detection pipeline and source classification are
presented in Pacaud et al. (2006). The 19 G pointings are part of
the XMDS (\textit{XMM} Medium Deep Survey, Chiappetti et al.
2005), which covers a total area of 3 deg$^{2}$ and which lies at
the heart of the XMM-LSS. About two thirds of the XMDS area are
covered in the optical band with the VVDS (VIRMOS VLT Deep Survey)
both by UBVRI photometry (Le F\`{e}vre et al. 2004) and by
multi-object spectroscopy with VIMOS (Le F\`{e}vre et al. 2005)
and by an associated radio survey at 1.4 GHz (Bondi et al. 2003).
Part of the VVDS area ($\sim$0.8 deg$^{2}$) is also covered by the
UKIRT Infrared Deep Sky Survey (UKIDSS, Dye et al. 2006; Lawrence
et al. 2007). Finally a large area of the XMM-LSS Survey has been
covered by 2dF observations in December 2003 as part of the 11
deg$^{2}$ medium sensitivity XMM Wide Angle Serendipitous Survey
(XWAS), using the 2dF optical multi fibre spectroscope on the AAT
(Tedds et al., in prep).

The present paper gathers a large and representative sample of 99
spectroscopically identified optical counterparts of X-ray point
sources selected in the [2-10] keV band. The goal of our paper is
to perform an internal comparison of the X-ray properties of our
sample with the properties of their optical counterparts. Firstly,
even if our sample is affected by selection effects (mostly
concerning the optical counterparts), it will be shown that the
conclusions of our work are not significantly affected by these
incompleteness issues. Our results will finally be compared with
the predictions of the standard AGN unified scheme (Antonucci
1993). Our work is a complementary analysis to the works of Tajer
et al. (2007) and Polletta et al. (2007). They have selected a
sample of 136 X-ray point sources detected at $\geq$3$\sigma$ in
the [2-10] keV band within a 1 deg$^{2}$ area of the XMDS. Their
goal was to probe the populations of AGN by a fit of SED templates
over their optical+MIR photometric data points in order to infer
the nature of the X-ray sources and their photometric redshifts.
They only performed an X-ray spectral analysis for 55 sources, as
most of them do not have enough counts.

The advantage of our study resides in that optical spectra with
secure identification and redshift are available for each of the
optical counterparts of the 99 point-like X-ray sources, and
second, our sample covers a much larger area, about 3 times larger
than the sample of Tajer et al. (2007), with about twice as many
extracted X-ray spectra. This allows us to draw some stronger and
safer statistical trends.

The sky area over which we have selected the AGN candidates has a
very large overlap with the 4.2 deg$^{2}$ used by Gandhi et al.
(2006) for an angular clustering analysis of AGN. While Gandhi et
al. (2006) have discussed the clustering properties of sub-samples
of absorbed and unabsorbed AGN separately, only minimal source
identification and classification criteria were adopted by them.
Our current paper is a big step towards fully classifying their
moderately bright sample of AGN candidates and eventually studying
the full three-dimensional clustering of AGN over a large sky
area.

Our paper is organized as follows : the X-ray sample and the
optical spectra are presented in Sect. 2. The optical
classification is discussed in Sect. 3. The X-ray data reduction
and the X-ray spectral analysis are presented in Sect. 4. In Sect.
5, we compare and discuss the optical obscuration and the X-ray
absorption within the framework of the AGN unified scheme. In
Sect. 6, we discuss the nature of the X-ray Bright Optically
Normal Galaxies, while in Sect. 7, we present 7 type 2 QSO
candidates. Finally, Sect. 8 provides a summary of the reported
results. Throughout this paper, we assume $H$$_{0}$= 70 km
s$^{-1}$ Mpc$^{-1}$, $\Omega$$_{\mathrm{m}}$=0.27 and
$\Omega$$_{\mathrm{\Lambda}}$=0.73, in accordance with the WMAP
cosmological parameters reported by Spergel et al. (2007).

\section{The samples}

\subsection{The sample of X-ray selected point-like sources}

We have used the most up-to-date X-ray catalog (Pierre et al.
2007) to define a sample of 612 X-ray selected point-like sources
which have both a log-likelihood of detection $>$20 (this roughly
corresponds to $>3\sigma$) in the [2-10] keV band (see Pacaud et
al. 2006 for full details), and a total number of counts$>$80 in
the [0.5-10] keV band. These targets are located within the whole
6 deg$^{2}$ of the XMM-LSS, which have been observed so far. We
used a minimum of 80 total counts in order to ensure sufficient
photon-statistics to perform an X-ray spectral analysis. Fig.
\ref{xmmlss} in the appendix shows this sample of 612 X-ray
sources, which are distributed among the 51 X-ray pointings.

As illustrated in the left panel of Fig. \ref{histofluxall} in the
appendix, most of the 612 X-ray sources have a 2-10 keV flux
between 8$\times$10$^{-15}$ and 8$\times$10$^{-14}$ erg s$^{-1}$
cm$^{-2}$, and the turnover of the flux distribution is around
10$^{-14}$ erg s$^{-1}$ cm$^{-2}$.

\subsection{The sample of $R\leq 22$ spectroscopically identified X-ray point
sources}

Presently, optical spectra with secure identification and
spectroscopic redshift are available for 99 X-ray sources out of
612. These 99 X-ray sources have been extracted in 26 X-ray
pointings, and have a limiting flux of 8$\times$10$^{-15}$ erg
s$^{-1}$ cm$^{-2}$ . Due to the fact that the X-ray pointings do
overlap, the corresponding total area is smaller, around 3
deg$^{2}$.

In order to get the optical spectrum of the counterpart of each
X-ray source, we have correlated our sample of 612 X-ray sources
with the catalog of optical spectra observed so far, out to
5$\arcsec$. Then, we only selected those X-ray sources which have
an optical spectrum with a secure identification and a
spectroscopic redshift. Proceeding this way, we end up with a sub
sample of 99 X-ray sources. The histogram of the separations
between the X-ray sources and the optical counterparts is shown in
the left panel of Fig. \ref{histodist} in the appendix. Finally,
we computed the probability that the association between an X-ray
source and its putative optical counterpart results from random
fluctuations using the following equation (Downes et al. 1986) :

\begin{equation}\label{proba}
p=1-exp(-\pi n(<m)r^{2}),
\end{equation}

where $r$ is the distance between the X-ray source and its optical
counterpart, and $n$($<$$m$) is the density of objects brighter
than the magnitude $m$ of the optical counterpart, which has been
obtained by counting the number of sources brighter than a given
$R$ magnitude based on the \textit{virphot} catalog. This
probability has been computed in order to check whether the
optical spectrum is really the one of the X-ray source. As in
Tajer et al. (2007), we ranked the probability that the
association itself is non random as "good" ($p<0.01$), "fair"
($0.01<p<0.03$) and "bad" ($p>0.03$). Out of the 94 optical
counterparts for which the $R$ band magnitude is available, 94
(100\%) are classified as "good" ones. This is expected because
most if not all of the optical counterparts are quite bright. This
thus allows us to be highly confident in the proposed association
between each X-ray source and its optical spectrum.

The optical spectra used in this work mainly issue from 3
spectroscopic runs: 79 of them have been taken with the 2dF in
December 2003 (as part of the XWAS project) in two overlapping
circular pointings, 2 degrees across and separated by 1.26
degrees. The pointing centers are :
$02^{\mathrm{h}}23^{\mathrm{m}}52^{s}$--$03^{\circ}49^{\arcmin}00^{\arcsec}$
and
$02^{\mathrm{h}}25^{\mathrm{m}}08^{\mathrm{s}}$--$05^{\circ}02^{\arcmin}27^{\arcsec}$,
respectively. The two pointings have been exposed for 4800 s and
3600 s respectively, both with the 300B grating. The achieved
spectral resolution of the 2dF spectra is around $R\sim600$ . The
S/N is about 5 at 5500 $\AA$ for a source of magnitude $V=21$.
Further details about the 2dF optical spectroscopy used in this
work are provided in the XWAS catalogue paper (Tedds et al., in
prep). Note that 3 optical sources observed with 2dF have also
been acquired independently with the Southern African Large
Telescope (SALT) in 2006 with the PG0300 ($R\sim400$) and the
PG0900 ($R\sim1200$) grisms during a PV phase. 9 spectra have been
obtained with FORS2 at the VLT during follow-up campaigns of the
XMM-LSS survey in 2002, 2003 and 2004. The combined exposure time
is 1 hour. 8 pointings have been exposed with the 600RI grism
($R\sim1000$) and one has been exposed with the 600z grism
($R\sim1400$). Finally, 11 spectra have been obtained with VIMOS
at the VLT within the VVDS (Le F\`{e}vre et al. 2005; Gavignaud et
al. 2006). These objects were observed with the low-resolution red
grism which covers the wavelength range 5500-9500 $\AA$ with a
spectral resolution $R\sim230$. The combined exposure time is 4.5
hours.

With this large number of spectroscopically identified X-ray
sources, we are in a good position to start an analysis in which
we characterize the population of X-ray sources in comparing their
X-ray and optical properties.

Before starting this analysis, we have checked whether the sub
sample of 99 X-ray sources constitutes a fair representation of
the whole sample, which includes 612 X-ray sources. To this end,
we first compared the [2-10] keV band flux distributions of these
two samples. The histogram presented in the left panel of Fig.
\ref{histofluxall} in the appendix shows the fraction of sources
as a function of the [2-10] keV flux. It can be seen that the
overall distributions of the flux for the sub sample of the 99
optically selected X-ray point sources and for the 513 still
optically unidentified X-ray sources are in a rather good
agreement with each other. Next, we also compared the hardness
ratios for these same two samples. The hardness ratio is defined
as :

\begin{equation}\label{hratio}
HR=\frac{H-S}{H+S},
\end{equation}

where $H$ is the background subtracted number of counts in the
hard [2-10] keV band and $S$ is the background subtracted number
of counts in the soft [0.5-2] keV band. The histograms of the
hardness ratio for the two samples are presented in the right
panel of Fig. \ref{histofluxall} in the appendix. This figure
shows the fraction of X-ray sources in each bin for the hardness
ratio. Once again, the two distributions are in good agreement
with each other.

From both panels of Fig. \ref{histofluxall} in the appendix, it
can be seen that the 99 spectroscopically identified X-ray sources
are slightly brighter and softer in the X-rays than the 513 still
optically unidentified X-ray point sources.

Furthermore we have performed a  Kolmogorov-Smirnov (K-S) test for
both the hardness ratio and the flux distributions between the 99
X-ray sources which have an optical spectrum and the remaining 513
X-ray sources which do not have an optical spectrum, in order to
test whether this trend is significant or not. We find that the
probability that the two samples are drawn from the same
population is $P=0.13$ and $P=0.15$, for the hardness ratio and
the X-ray flux, respectively. Therefore, our sample of 99
optically identified X-ray point sources, for which the
determining criterion to be included in the sample is the optical
spectrum (which is not a uniform criterion at all) statistically
constitutes a fair representation of the whole sample of the 612
X-ray sources. Optical selection effects will be discussed in the
next section.

\section{The Optical classification criteria}

To differentiate between type 1 and type 2 AGN, a value between
1000 km s$^{-1}$ and 2000 km s$^{-1}$ is often adopted for the
relevant Full Width at Half Maximum (FWHM) border value of their
corresponding emission lines (e.g. Page et al. 2006a; Caccianiga
et al. 2004). In the present work, we did not use 1000 km
s$^{-1}$, as the FWHM of some well known forbidden emission lines
(e.g. [OII]) amounted up to FWHM$\sim$1100 km s$^{-1}$. We decided
to choose 1500 km s$^{-1}$ as the dividing value to classify an
AGN as either a type 1 or a type 2 object.

The FWHMs of all emission lines have been directly measured by
fitting gaussian profiles, for the 99 optical spectra, using the
IRAF\footnote{\textsc{IRAF}, which is an Image Reduction and
Analysis Facility, is distributed by the National Optical
Astronomy Observatories.} task \texttt{splot} from the package
\texttt{onedspec}. We have checked that the spectral resolution of
the optical spectra (the lowest being $R\sim230$) allows us to
distinguish a narrow emission line from a broad emission line,
using 1500 km s$^{-1}$ as the dividing value for the FWHM.\\
Following our convention, we have divided our sample into two
optical classes :

- \textbf{Type 1 objects}, including

$\bullet$ 61 sources showing broad ($V$$_{\mathrm{FWHM}}$ $\geq$
1500 km s$^{-1}$) permitted emission lines. These are Broad
Emission Line AGN (BLAGN).

- \textbf{Type 2 objects}, including

$\bullet$ 35 sources showing narrow ($V$$_{\mathrm{FWHM}}$ $<$
1500 km s$^{-1}$) permitted emission lines. These are Narrow
Emission Line Galaxies (NELG).

$\bullet$ 3 sources showing no emission lines in their spectra.
These are Absorption Line Galaxies (ALG).

\vspace{0.5cm}

These values are gathered in Table 1. The fraction of type 1
(BLAGN) and type 2 objects (NELG+ALG) is thus 62\% and 38\%
respectively.

\begin{table}
\begin{tabular}{|c|c|c|c|}
  \hline
  Class & Spectral type & Number  \\
  \hline
  \textbf{Type 1} & BLAGN & \textbf{61} \\
  \hline\hline
  \textbf{Type 2} & NELG & 35 \\
   & ALG & 3  \\
   \hline
   & NELG+ALG & \textbf{38} \\
   \hline\hline
 \textbf{Total } &  & \textbf{99}  \\ \hline
\end{tabular}
\caption{Optical spectroscopic classification.}
\label{tableagnopt}
\end{table}

\begin{figure}[ht!]
  \includegraphics[angle=0,width=8cm]{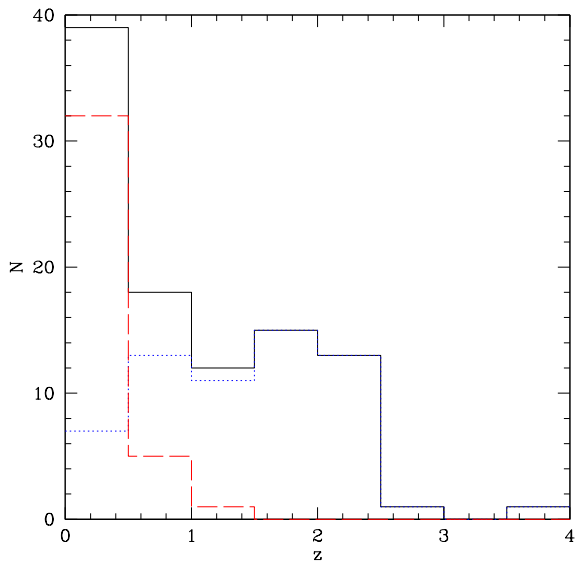}
  \caption{Spectroscopic redshift distribution of the 99 optical counterparts in our
  sample. The solid histogram refers to the whole sample, the long-dashed histogram
  refers to type 2 AGN and the dotted histogram refers to type 1 AGN.}  \label{histozspec}
\end{figure}

Concerning the optical selection effects, we are biased against
faint (typically $R>22$) optical counterparts, as can be seen in
the right panel of Fig. \ref{histodist} in the appendix. In order
to test whether the optical selection effects will significantly
hamper the comparison between the X-ray and the optical
properties, we searched for the fraction of similar types of AGN
in much deeper surveys. Eckart et al. (2006), who have identified
spectra of optical counterparts down to $R=24$, have found that
51\% of their optical counterparts are BLAGN, which represents a
fraction of BLAGN which is not significantly different from ours
(within 2 $\sigma$).

If we restrict their sample down to $R<22$, which is approximately
the limiting magnitude of our analysis, they have identified 63\%
of BLAGN, which is very consistent with our corresponding fraction
of BLAGN (within 1 $\sigma$).

All this suggests that our internal comparison between the X-ray
and the optical properties does not suffer much from
incompleteness issues, at least on statistical grounds.

Note that the fraction of BLAGN, NELG and ALG in our sample is
also consistent with several other published studies (e.g. Fiore
et al. 2003, Silverman et al. 2005). Della Ceca et al. (2004) have
defined two samples : one in the [0.5-4.5] keV band, which is the
XMM-Newton Bright Source Sample (BSS, hereafter) and the other one
in the [4.5-7.5] keV band, which is the XMM-Newton Hard Bright
Source sample (HBSS, hereafter). Concerning the BSS, they find a
significantly higher fraction of broad line AGN (around 85\%).
This can be explained by the fact that it corresponds to a band
which is softer than our selected [2-10] keV band. In the HBSS,
they find a fraction which is more consistent with ours (around 70
\%, within 1 $\sigma$).

The right panel of Fig. \ref{histodist} in the appendix shows our
own $R$ band magnitude histogram of the 94 identified sources for
which an $R$ band magnitude is available. The $R$ band magnitudes
come from the UKST plate scans from the supercosmos archives in
Edinburgh (equivalent to Cousins Vega magnitudes) and from the
VVDS. Note that the lack of the $R$ band magnitude for the
remaining 5 optical counterparts is solely due to partial optical
imaging coverage.

Most of the sources have an $R$ band magnitude between 17 and 21.
We have mostly identified bright optical counterparts, due to the
fact that most of our optical spectra come from the shallow 2dF
run.

As shown in Fig. \ref{histozspec}, type 1 AGN in our sample have
been detected out to $z\sim4$. These type 1 AGN have a broad
distribution in redshift, most of them having $z<2.5$, which is
consistent with the one observed by Tajer et al. (2007); see also
Silverman et al. (2005). We have mostly identified type 2 AGN at
$z<0.6$. The distribution of type 2 AGN in our sample drops at a
slightly lower redshift value than the one observed by Tajer et
al. (2007). However this can mostly be accounted for  by the fact
that the bulk of our identified optical counterparts have a
brighter optical magnitude. And thus statistically, they should
lie at a lower redshift.

\section{X-ray spectral analysis}

\subsection{X-ray data reduction}

We have extracted the X-ray spectra for each of the 99 X-ray point
sources of our sample, detected in the [2-10] keV band, for which
we have an accurate redshift. Note that each X-ray source detected
in the XMM-LSS survey has an off-axis distance smaller than
13$\arcmin$, as the X-ray pipeline (Pacaud et al. 2006) detects
X-ray sources with an off-axis distance up to 13$\arcmin$.

The X-ray data reduction has been performed using the most recent
version of the XMM-Newton Science Analysis System (\textsc{SAS}
v7.0.0). All valid event patterns (PATTERN 0-12) were used in
constructing the MOS spectra. For the pn spectra, only the single
and double events (PATTERN 0-4) have been used. Events were
extracted for each source using the \textsc{SAS} task
\texttt{evselect} in a circular region with a radius varying
between 20$\arcsec$ and 30$\arcsec$, depending on the off-axis
distance of the X-ray source. The background events have been
extracted in the nearest source free circular region, on the same
CCD chip, excluding areas near CCD gaps. We have tried several
sizes for the background extraction region, between 20$\arcsec$
and 60$\arcsec$ : a small region is not really representative of
the background underneath the source (noise problem) whereas a
large region includes background counts too far away from the
source where the background might be significantly different.
Moreover, it would also increase the probability to include other
X-ray sources in the background region. We ended up choosing a
radius around 40$\arcsec$ for the background extraction region as
we found it to be a good compromise. The background regions have
been chosen manually, in order to avoid to include other X-ray
sources.

In order to perform a proper spectral analysis, we created the
redistribution matrix file (RMF) and the ancillary response file
(ARF) for each X-ray source and for each  of the used detectors,
using the SAS tasks \texttt{rmfgen} and \texttt{arfgen},
respectively. To make full use of the available X-ray data, we
have simultaneously fitted the X-ray spectra of the 3 detectors
(MOS1, MOS2, and pn), whenever possible.

For the cases consisting of X-ray sources detected over two X-ray
pointings, for which the off-axis distances are less than
12$\arcmin$ and the separation on the two X-ray pointings is less
than 1$\arcmin$, we have jointly extracted their X-ray spectra, in
order to keep a good PSF and to have rmf files that are close
enough to each other. There were 5 such cases. For the other
cases, we have chosen the X-ray pointing with the longest exposure
time, while if the exposure times are equivalent, we have chosen
the pointing for which the X-ray source has the smaller off-axis
distance.

\subsection{Results}

The X-ray spectra have been analyzed using \textsc{XSPEC}
(v11.3.0). We have kept the [0.3-10] keV region for spectral
fitting and we ignored bad energy channels. We fitted as many of
the 3 detectors as possible. We binned the spectra with at least 8
energy bins, in order to have enough energy bins to fit a powerlaw
model to the X-ray spectra. When at least 15 counts are available
in each energy bin, $\chi$$^{2}$ statistic was used. Otherwise, we
used Cash statistic (Cash 1979), in which case we binned the
spectra to have at least 5-10 counts in each energy bin. Among our
99 X-ray spectra, 41 of them have been fitted using $\chi$$^{2}$
statistic and 58 of them have been fitted using Cash statistic.
When the $\chi$$^{2}$ statistic was applied, we verified that the
Cash statistic yielded consistent results.

For each X-ray spectrum, we first fitted the spectrum using an
absorbed powerlaw model with galactic hydrogen column density
along the line of sight ($N$$_{\mathrm{H}}$=2.6 10$^{20}$
cm$^{-2}$; Dickey \& Lockman 1990) and a possible intrinsic
absorption component at the source redshift. This corresponds to
the \textsc{XSPEC} model phabs$\ast$zphabs$\ast$pow, fixing the
galactic column density to the value given above and using the
spectroscopic redshift of the source. When $\chi$$^{2}$ statistic
was used, we fitted the spectrum setting both the intrinsic column
density ($N_{\mathrm{H}}^{\mathrm{int}}$) and the photon index
($\Gamma$) as free parameters. In cases where the fitted value of
$N_{\mathrm{H}}^{\mathrm{int}}$ is consistent with 0 (at the 95\%
confidence level), we refitted the X-ray spectrum using the
\textsc{XSPEC} model phabs$\ast$zpow, with $\Gamma$ as the only
free parameter.

Otherwise, when Cash statistic was used, we never set both the
$N_{\mathrm{H}}^{\mathrm{int}}$ and the $\Gamma$ as free
parameters, as there were not sufficient counts to constrain both
parameters. So, when using Cash statistic, the fit was first
performed with $N_{\mathrm{H}}^{\mathrm{int}}$ as a free
parameter, fixing $\Gamma$ to a value of 1.9, which is a value
representative of broad line AGN (Turner \& Pounds 1989; Nandra \&
Pounds 1994). Each time the inferred
$N_{\mathrm{H}}^{\mathrm{int}}$ was consistent with 0 (at the 95\%
confidence level), we refitted the spectra only taking into
account the galactic absorption component (\textsc{XSPEC} model
phabs$\ast$zpow) and with $\Gamma$ as the only free parameter. 7
sources out of 99 are better fitted (significant increase of the
goodness of fit) when a more complicated model than a powerlaw is
used : 1 needed an absorbed powerlaw+black body model, 1 two power
laws, one absorbed and one unabsorbed with the same value for
$\Gamma$, and for 5 sources, the presence of an emission line was
required (2 have been tentatively identified as the Fe K line at
6.4 keV, using the spectroscopic redshift of the source).

We have classified our X-ray sources using a discriminating value
of $N_{\mathrm{H}}^{\mathrm{int}}$= 10$^{22}$ cm$^{-2}$ for the
best fit value of the intrinsic hydrogen column density. This
value has been chosen as it corresponds to the column density of
neutral hydrogen needed to hide the broad line regions for clouds
which have a standard gas-to-dust ratio (Silverman et al. 2005).

Above this threshold, the AGN is considered as absorbed in the
X-rays (hereafter type II). Otherwise, the AGN is unabsorbed
(hereafter type I). Using this criterion we end up with 79 X-ray
sources classified as type I AGN and 20 sources classified as type
II AGN. Tajer et al. (2007) report a significantly higher fraction
of type II AGN (49\%) but this is mostly due to our different
selection method. Indeed, as our study is only keeping X-ray
sources with more than 80 counts in the [0.5-10] keV band, it is
likely that we identify mostly type I AGN.

Furthermore, as Tajer et al. (2007) use photometric redshifts,
they go much deeper in magnitude (although with larger redshift
uncertainties) for the optical photometric points over which a
spectral energy distribution template is fitted in order to assess
both the redshift and identify the nature of the X-ray source.

The observed frame X-ray flux, and the rest frame, deabsorbed
(intrinsic) X-ray luminosity in the [2-10] keV band have been
derived for the 99 X-ray sources, directly from the model fitted
to the X-ray spectrum.

\begin{figure}
  \includegraphics[angle=0,width=8cm]{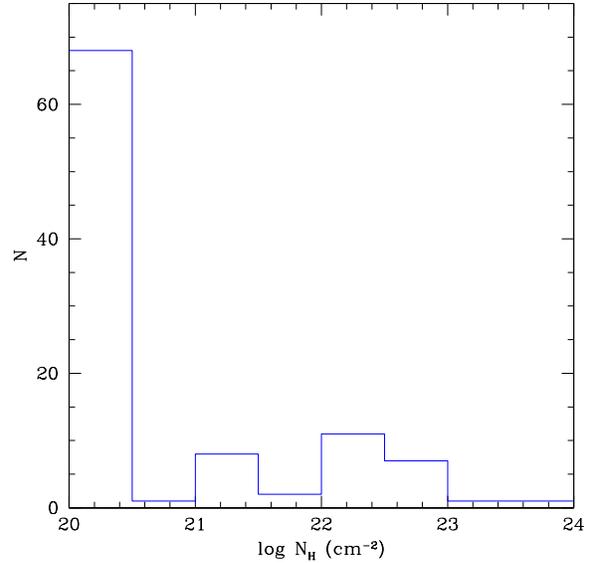}
  \caption{Column density distribution (galactic+intrinsic component) for the whole sample.}
  \label{histonhintall}
\end{figure}

The distribution of $N_{\mathrm{H}}$ for our sample is shown in
Fig. \ref{histonhintall}. The first bin in the histogram
corresponds to the fixed galactic value for the hydrogen column
density. This refers to X-ray spectra for which a power law model
with no intrinsic absorption has been fitted, as we have checked
that X-ray absorption is inconsistent at the 95\% confidence range
for these X-ray sources.

\begin{figure}
  \includegraphics[angle=0,width=8cm]{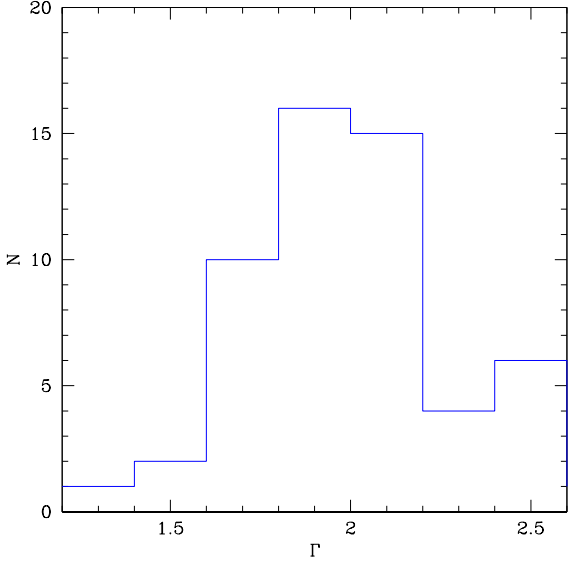}
  \caption{Distribution of the photon indices that have been fitted for the X-ray sources
  classified as type I.}
  \label{histogammall}
\end{figure}

Concerning the type I X-ray sources we have made the histogram of
the 60 fitted values of $\Gamma$ shown in Fig. \ref{histogammall}.
Our distribution of the fitted value for the photon index has a
mean value of $\Gamma=2.01\pm0.28$, which is consistent with the
works of Page et al. (2006a), Caccianiga et al. (2004), Mainieri
et al. (2002, 2006), and Mateos et al. (2005).

\section{Optical obscuration versus X-ray absorption}

In this section, we compare optical obscuration and X-ray
absorption within the framework of the AGN unified scheme. The
simple orientation-based unified model (Antonucci 1993) predicts a
strict correlation between the optical obscuration and the X-ray
absorption. However, recently, quite a few studies have claimed to
find objects for which the X-ray and the optical classifications
do not match. For example, Panessa \& Bassani (2002) have
presented a significant number of type 2 AGN whose X-ray spectra
are indicative of a mild (or absent) absorption. They tentatively
estimated this percentage in the range 10\%-30\%. Another
important exception to the X-ray/optical classification
correlation is the discovery of type 1 AGN with large X-ray
absorption (e.g Fiore et al. 1999), some of them being Compton
thick (Gallagher et al. 2006). Finally, very recently, Page et al.
(2006a) also reported a significant number of type 2 AGN with low
X-ray absorption.

\begin{figure}
  \includegraphics[angle=0,width=8cm]{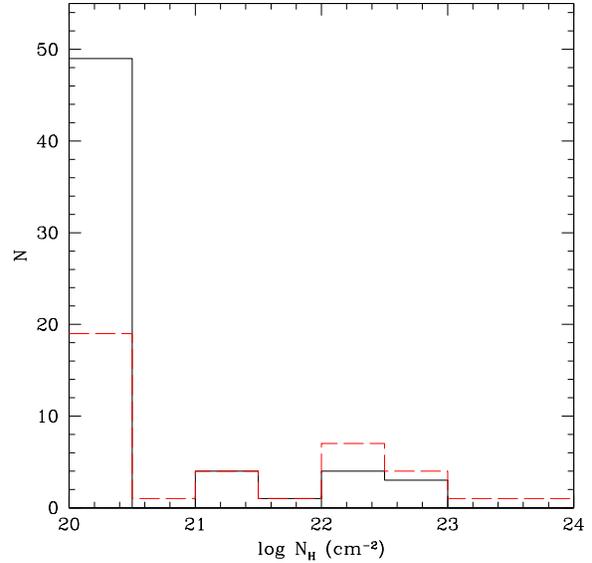}
  \caption{Column density distribution (galactic+intrinsic component)
  for type 1 AGN (solid histogram)
  and for type 2 AGN (long dashed histogram).}
  \label{histonhint}
\end{figure}

In the present work, we also find a significant number of X-ray
sources for which the X-ray and the optical classifications do not
match. Fig. \ref{histonhint} illustrates the column density for
type 1 AGN (solid histogram) and for type 2 AGN (long dashed
histogram). This figure shows that 20\% of the type 1 objects have
$N_{\mathrm{H}}$$>$ 10$^{21}$ cm$^{-2}$ and 11\% have
$N_{\mathrm{H}}$$>$10$^{22}$ cm$^{-2}$. Tajer et al. (2007) find
that about 31\% of their type 1 AGN show strong X-ray absorption
($N_{\mathrm{H}}$$>$10$^{22}$ cm$^{-2}$). They inferred the
intrinsic column density from a sample of X-ray spectra and from
an analysis of the hardness ratios for the X-ray sources which do
not have a sufficient number of counts to extract their X-ray
spectra. Only taking into account the X-ray sources for which an
X-ray spectrum has been extracted, they find a value much closer
to ours, around 9\% which is in good agrement.

On the other hand, 66\% of our type 2 objects have
$N_{\mathrm{H}}$$<$10$^{22}$ cm$^{-2}$ and 53\% have
$N_{\mathrm{H}}$$<$10$^{21}$ cm$^{-2}$. This is consistent with
the results of Page et al. (2006a) who find that about 68\% of
their type 2 AGN are showing very low absorption in the X-rays.

\begin{table}
\begin{tabular}{|c|c|c|c|}
\hline
   & Type I & Type II & N\\
  \hline
  \textbf{Type 1} & 54 & 7 & 61 \\
  \textbf{Type 2} & 25 & 13 & 38 \\ \hline
    N & 79  &  20 & \textbf{99} \\
    \hline
\end{tabular}
\caption{Number of sources as a function of the optical (Type 1 or
Type 2) and the X-ray (Type I or Type II) classifications.}
\label{tableagn}
\end{table}

Table 2 shows the number of sources divided according to their
X-ray and optical classifications. This table shows that there is
a large probability that a BLAGN is a type I, as 54$/$61 (89\%) of
them are unabsorbed in the X-rays. Also, there is a large
probability that an absorbed AGN is a type 2, as 13$/$20 (65\%) of
them only present narrow emission lines in their optical spectra.
This trend has been previously observed many times (e.g Mainieri
et al. 2006; Page et al. 2006a; Tajer et al. 2007). However, in
overall, the optical and the X-ray classifications are only
matched for 68\% of the AGN in our sample. At first sight, 68\%
still seems to be large. In order to test in more details this
correlation, we thus computed the fourfold point correlation
coefficient (hereafter r) which is defined as :

\begin{equation}\label{correl}
r=\frac{n_{11}n_{22}-n_{12}n_{21}}{\sqrt{n_{1.}n_{2.}n_{.1}n_{.2}}},
\end{equation}
where $n_{ij}$ is the value of the element $ij$ of the table;
$n_{1.}$ corresponds to $n_{11}$+$n_{12}$; $n_{2.}$ corresponds to
$n_{21}$+$n_{22}$; $n_{.1}$ corresponds to $n_{11}$+$n_{21}$ and
$n_{.2}$ corresponds to $n_{12}$+$n_{22}$. The fourfold point
correlation coefficient is similar to the ordinary correlation
coefficient : when this coefficient is equal to 1, there is a
strict correlation, when it is equal to 0, there is no correlation
at all, and finally when it is equal to $-$1, there is an
anti-correlation. We introduce this correlation coefficient
because it corresponds to a more rigorous way to quantify the
correlation between the X-ray and the optical classifications : if
no correlation is observed between the X-ray and the optical
classifications ($r=0$), we would find 25\% (in case of
equipartition among the two types) of the objects in each box of
Table 2., which would mean that the X-ray and the optical
classification would be matched for up to 50\% of the X-ray
sources.

\begin{figure}
  \includegraphics[angle=0,width=8cm]{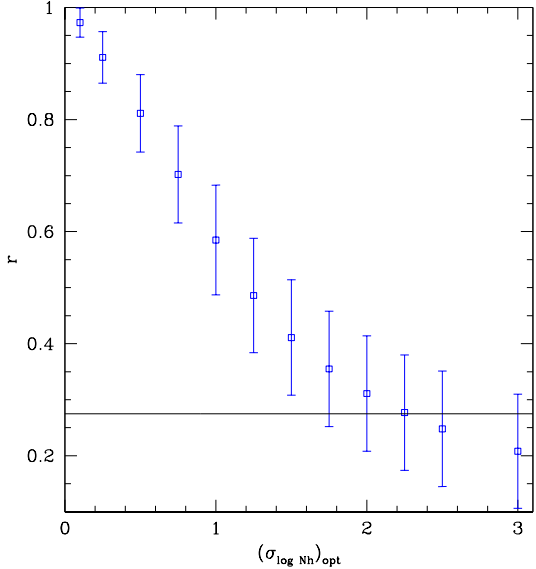}
  \caption{The average value of the $r$ coefficient over 100,000 simulations as a function
  of the standard deviation $\sigma$ of the assumed gaussian distribution of the logarithm of the column density
  in the optical. The error bars for the $r$ value correspond to 1 $\sigma$. The solid
  line corresponds to the value $r=0.275$ that we obtain for our sample of 99 X-ray sources.}
  \label{rsimu}
\end{figure}

We thus applied eq. (\ref{correl}) to the elements of Table
\ref{tableagn} and found a value of $r=0.275^{+0.132}_{-0.098}$.
We defined one-sided error bars, which have been computed assuming
poisson errors according to Gehrels (1986) for low number
statistic, on each of the number of sources $n_{ij}$ and using the
error propagation equation starting from eq. (\ref{correl}). We
have tested these error bars in producing 1,000,000 Monte Carlo
simulations assuming Poisson statistic in the presence of
correlation. We find very similar error bars, even if slightly
smaller. Finally, we have produced the same kind of simulation but
in the presence of no correlation (i.e. $r=0$). The dispersion
obtained around the value $r=0$ is 0.1. Thus, the value of $r$
that we obtain for the whole sample ($r=0.275^{+0.132}_{-0.098}$)
is consistent with a mild correlation but is significantly
(2$\sigma$ level) different from the case $r=0$. Note that the
value of $r$ that we have obtained is totally inconsistent with a
strict correlation between the X-ray and the optical
classifications, even if the two classifications are matched for
up to 68\% of the X-ray sources in the sample, as a strict
correlation between the X-ray and the optical classification would
yield a value compatible with $r=1$.

In order to better clarify the meaning of the result $r=0.275$, we
have produced 100,000 samples of 99 X-ray sources for which we
have computed the coefficient $r$, proceeding in the following
way: in each simulated sample, we have assumed for the optical
obscuring column in each source a value drawn from a gaussian
distribution around the $N_{\mathrm{H}}$ derived from the X-ray
spectral analysis. We have then classified the X-ray sources using
a threshold of $N_{\mathrm{H}}$$=$10$^{22}$ cm$^{-2}$, both in the
X-rays (actually no change occurs) and in the optical. Finally, we
have computed the average value of the coefficient $r$ over the
100,000 simulations, as a function of the standard deviation
$\sigma$ of the gaussian distribution of the logarithm of the
column density in the optical. The results that we have obtained
from these simulations are shown in Fig. \ref{rsimu}. This Figure
clearly indicates that a value of $r=0.275$ corresponds to a
rather large width for the gaussian distribution of the logarithm
of the column density in the optical, which means that there is at
most a mild correlation between the X-ray and the optical
classifications. More quantitatively, these simulations have shown
that a value of $r=0.275$ is consistent with the $\sigma$ of the
gaussian distribution being equal to 2. In other words, it means
that 32\% of the X-ray sources have their column density in the
X-rays and in the optical which differ by more than a factor 100.
In order to take into account the fact that X-ray sources with a
column density fixed to the galactic value might have mild
intrinsic X-ray absorption, we have obtained new simulations
setting $N$$_{\mathrm{H}}$=10$^{21}$ cm$^{-2}$ for these X-ray
sources. With this value, we find that 32\% of the X-ray sources
have their column density in the X-rays and in the optical which
differ by more than a factor 40. Therefore, these simulations have
clearly shown that there is at most a mild correlation between the
X-ray and the optical absorption properties.

As it will be shown later, sources \#44 and \#63 (both being
unabsorbed X-ray sources, classified as type 2 AGN) are embedded
in an X-ray galaxy cluster. So for these 2 sources, the difference
between the X-ray and the optical classifications might be due to
a soft contribution from the X-ray galaxy cluster. We have thus
also computed $r$ excluding these two X-ray sources. We end up
with $r=0.294^{+0.134}_{-0.100}$. Even if slightly higher, the
conclusion remains unchanged.

When applied to the sample of Tajer et al. (2007), we find a value
of $r=0.252^{+0.106}_{-0.085}$. Their result is consistent with
ours within the error bars. Their value of $r$ is also consistent
with the absence of a strict correlation between the X-ray and the
optical classifications.

We then computed the fourfold point correlation coefficient $r$ as
a function of the X-ray flux and the X-ray luminosity in the
[2-10] keV band : we first defined 3 sub samples having a flux
limit of 1, 3 and 5$\times$10$^{-14}$ erg s$^{-1}$ cm$^{-2}$. The
$r$ values obtained are $0.228^{+0.140}_{-0.103}$,
$0.163^{+0.252}_{-0.157}$ and $0.433^{+0.431}_{-0.220}$,
respectively.

We then also defined 3 sub samples having a luminosity larger than
10$^{42}$ erg s$^{-1}$, 10$^{43}$ erg s$^{-1}$ and 10$^{43.5}$ erg
s$^{-1}$. We end up with $r$ values of $0.285^{+0.138}_{-0.102}$,
$0.205^{+0.191}_{-0.127}$ and $0.261^{+0.331}_{-0.166}$,
respectively.

\begin{figure}
  \includegraphics[angle=0,width=8cm]{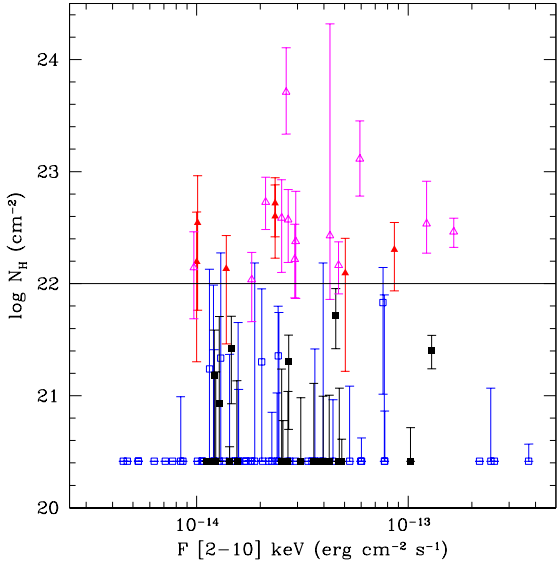}
  \caption{Column density distribution (galactic+intrinsic
  component) as a function of the absorbed [2-10] keV flux for the
  99 X-ray sources. Squares are X-ray sources classified as type I
  while triangles correspond to X-ray sources classified as type II.
  Filled symbols are the sources for which the
  X-ray and the optical classifications do not match. The error bars correspond to the 95\%
  confidence interval. Data points without error bars correspond to sources for which the
  column density has been fixed to the galactic value. These
  column densities are consistent with the galactic value and the presence of
  absorption in the relevant X-ray spectra is at least rejected at the 95\% level.
  The horizontal line corresponds to the dividing line between
  type I X-ray sources (for which $N_{\mathrm{H}}$$<$ 10$^{22}$ cm$^{-2}$) and
  type II X-ray sources ($N_{\mathrm{H}}$$>$ 10$^{22}$ cm$^{-2}$).}
  \label{fluxnh}
\end{figure}

Fig. \ref{fluxnh} shows the hydrogen column density as a function
of the absorbed [2-10] keV band flux. This figure clearly shows
that the fraction of sources for which the X-ray and the optical
classifications do not match does not seem to increase
significantly towards faint X-ray fluxes, i.e. there is no bias or
selection effect in flux. Note that 16 X-ray sources out of 99
have their 95\% confidence intervals crossing the discriminating
line of $N_{\mathrm{H}}$$=$10$^{22}$ cm$^{-2}$ (11 X-ray absorbed
and 5 X-ray unabsorbed sources). In order to estimate their
influence on the fourfold point correlation coefficient $r$, we
switched their X-ray classifications (11 X-ray unabsorbed and 5
X-ray absorbed) and have computed the corresponding correlation
coefficient $r$. We obtain a value of $r=0.097^{+0.144}_{-0.097}$.
We have also computed the correlation coefficient excluding these
16 X-ray sources: $r=0.281^{+0.161}_{-0.098}$.

Therefore the influence of these 16 ambiguous X-ray sources does
not significantly alter our results on the comparison between the
X-ray and the optical properties, e.g., at most a mild correlation
is observed.

Therefore, the mismatch we observe between the X-ray and the
optical classifications is a general result which does not
strongly depend on the X-ray flux or luminosity. Note that for a
flux greater than 5$\times$10$^{-14}$ erg s$^{-1}$ cm$^{-2}$ or a
luminosity larger than 10$^{43.5}$ erg s$^{-1}$, this result is
not as significant anymore. We obtain a value of
$r=0.348^{+0.151}_{-0.101}$ when considering the sample of Page et
al. (2006a), which is consistent with our result.

Silverman et al. (2005) find that 81\% of their X-ray sources can
be easily interpreted in the context of current AGN unification
models. Proceeding the same way, we obtain a value of at most
$r=0.58$ for their sample: it has to be considered as an upper
limit: as most of their inferred column density are upper limits,
we cannot tell whether the X-ray and the optical classifications
are matched for those cases. Their value, even if larger, is still
consistent with our results, within the error bars. Very recently,
Mainieri et al. (2006) have presented a sample of 135 X-ray
sources in the XMM-COSMOS field. From their data, we infer a
coefficient $r=0.458^{+0.103}_{-0.079}$. Once more, this value of
$r$ is consistent with only a mild correlation between the X-ray
and the optical properties, in good agreement with our own
results.

\begin{figure}[ht!]
  \includegraphics[angle=0,width=8cm]{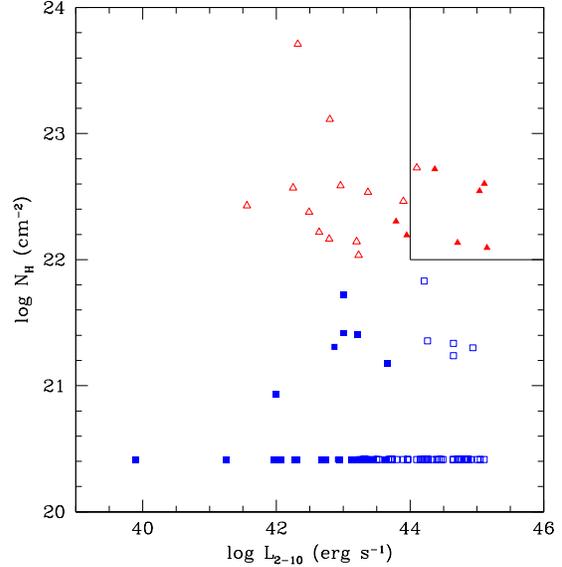}
  \caption{Column density (galactic+intrinsic component) as a function of
  the 2-10 keV intrinsic rest-frame luminosity. Squares are X-ray sources classified as type I
  while triangles are those classified as type II.
  Filled symbols are the sources for which the
  X-ray and the optical classifications do not match. The solid lines mark the region
  where type II QSOs should be found.}
  \label{lumnh}
\end{figure}

Finally, our results are significantly different from those of
Caccianiga et al. (2004) who have defined a sample of 28 bright
X-ray sources with spectroscopic identifications: they find that
the strict correlation predicted by the unified scheme is almost
always respected : all their type 1 objects are only showing mild
absorption in the X-rays. On the contrary, all but one of their
type 2 are characterized by column densities larger than 10$^{22}$
cm$^{-2}$. So 96\% of their X-ray sources have a similar
classification in the X-rays and the optical. We have computed the
fourfold point correlation coefficient for their analysis. We find
a value of $r$=0.91$\pm$0.26, showing that there is a strong
correlation ($r=1$) between the X-ray and the optical
classifications for the X-ray sources in their sample, which is
totally consistent with the unified scheme. The significant
differences with our sample are that their sample is solely
composed of X-ray sources with very bright fluxes
($F$$_{2-10}$$>$8$\times10$$^{-14}$ erg s$^{-1}$ cm$^{-2}$) and
that their sample is flux limited in the [4.5-7.5] keV band.

If we restrict our sample to X-ray sources brighter than this flux
limit, we only end up with 9 X-ray sources, for which the
coefficient $r$=$0.316^{+0.645}_{-0.282}$. So, at these very
bright X-ray fluxes, our results are consistent within the error
bars, with those of Caccianiga et al. (2004). However, this is
mainly due to the lack of identification of very bright X-ray
sources in our sample : only 9 X-ray sources out of 99 have
$F$$_{2-10}$$>$8$\times$10$^{-14}$ erg s$^{-1}$ cm$^{-2}$. So, it
would be very interesting to increase the size of our sample,
especially towards very bright X-ray fluxes, in order to check
whether the absence of a correlation between the X-ray and the
optical classifications is still observed.

In Fig. \ref{lumnh}, we have plotted the column density as a
function of the 2-10 keV intrinsic rest-frame luminosity. No
correlation between the absorbing column and the intrinsic X-ray
luminosity is found.

In this section, we have shown that, statistically, no strong
correlation is observed between the X-ray and the optical
properties. This result seems to contradict the predictions of the
AGN unified scheme. Therefore, in the following two subsections,
we will now discuss the physical nature of these 32 X-ray sources
for which the X-ray and the optical classifications do not match,
in order to investigate in further details whether the predictions
of the standard unified scheme are really not met. We will discuss
the 25 unabsorbed X-ray sources presenting only narrow emission
lines in their optical spectra in Sect. 5.1, and the 7 absorbed
X-ray sources presenting obvious and broad emission lines in their
optical spectra, in Sect. 5.2.

\begin{figure*}[ht!]
\centering
  \includegraphics*[angle=270,width=7.9cm,bb=573 43 77 721]{7778fig8a.ps}
  \includegraphics*[angle=0,width=9cm]{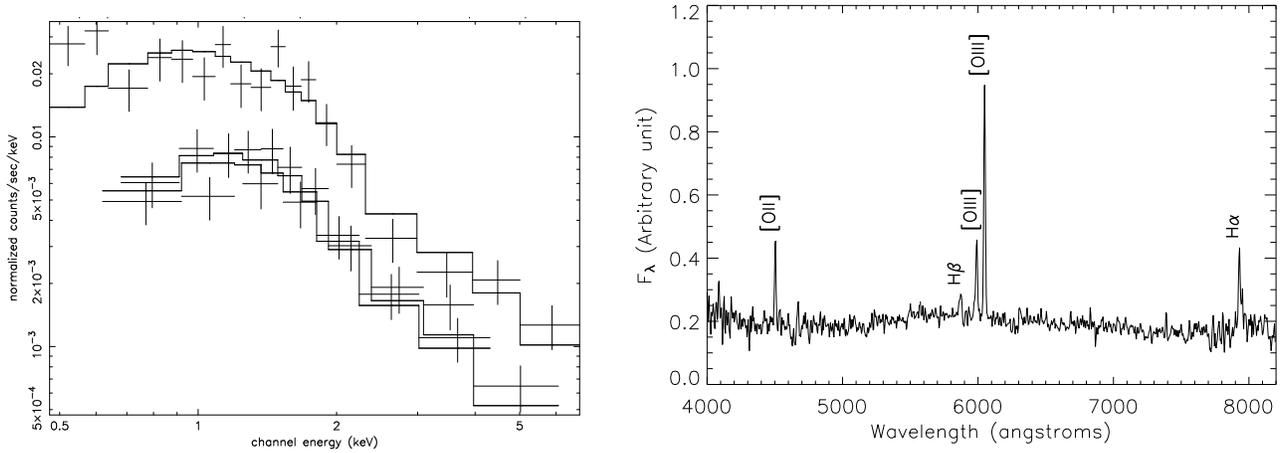}
  \caption{MOS1 MOS2 (the two lower functions) and pn (upper one) X-ray spectra of a source which is only showing
  mild X-ray absorption with $N_{\mathrm{H}}^{\mathrm{int}}$$=$2.5 10$^{21}$ cm$^{-2}$.
  The crosses represent the data points and the solid
  lines represent the folded model simultaneously fitted to the data of the 3 detectors (left).
  The optical counterpart has been identified as a Seyfert 2 lying at a redshift $z=0.207$. We clearly observe the
  [OIII] lines around 6000 $\AA$ (right). This is source \#53 in Table 4.}  \label{class12}
\end{figure*}

\subsection{Unabsorbed AGN lacking broad emission lines in their
optical spectra}

\subsubsection{Broad band properties}

Fig. \ref{class12} shows a representative example of an unabsorbed
X-ray source, which is classified as a type 2. This X-ray source
is showing mild absorption
($N_{\mathrm{H}}$$=$2.5$\times$10$^{21}$ cm$^{-2}$) with $\Gamma$
fixed to 1.9. This is source \#53 in Table 4, which lists the
X-ray and optical properties of all the 99 X-ray sources. We have
found 25 such cases amongst our sample.

\begin{figure}
  \includegraphics[angle=0,width=8cm]{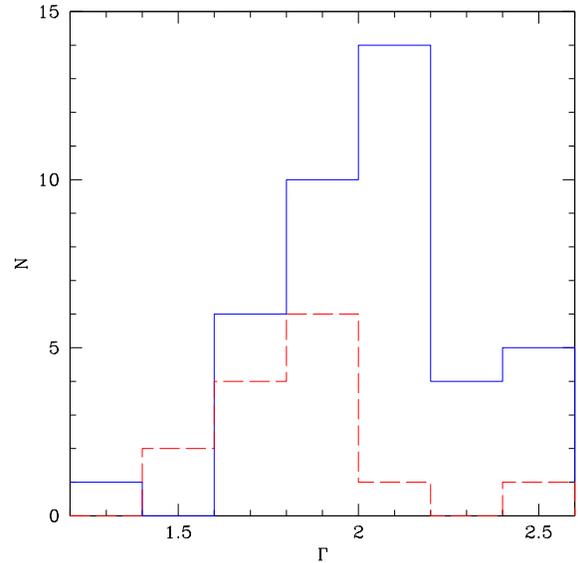}
  \caption{Distribution of the photon index for type I X-ray sources
  which are classified as type 1 (solid histogram) and as type 2 (long dashed histogram).}
  \label{histogamma}
\end{figure}

We have compared the distribution of the photon index, for the
unabsorbed X-ray sources which are type 1, and type 2 (Fig.
\ref{histogamma}). The distribution of the photon index for the
type 2 is significantly lower than the one for the type 1. The
distribution for the type 2 is peaking around $\Gamma\sim1.9$.

\begin{figure}
  \includegraphics[angle=0,width=8cm]{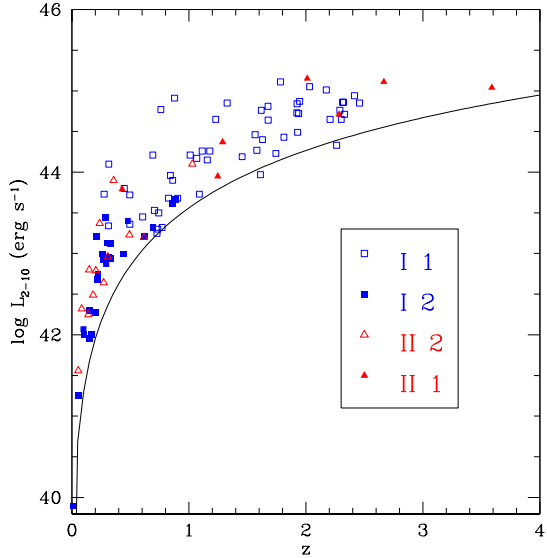}
  \caption{2-10 keV intrinsic rest-frame luminosities as a function of
  redshift for our sample of X-ray sources. Squares are X-ray sources classified as type I
  while triangles are those classified as type II.
  Filled symbols are the sources for which the
  X-ray and the optical classifications do not match.
  The solid line is showing the 2-10 keV luminosity as a function of
  redshift for a 2-10 keV limiting flux of 8$\times$10$^{-15}$ erg
  s$^{-1}$ cm$^{-2}$.}
  \label{lumint}
\end{figure}

As can be seen in Fig. \ref{lumint}, the vast majority of these
discrepant objects (filled squares; 84\%) are lying below $z=0.5$.
However they span a broad range of X-ray luminosities, from
$L$$_{2-10}$$=$10$^{40}$ erg s$^{-1}$ to almost $L$$_{2-10}$$=$5
10$^{43}$ erg s$^{-1}$.

As a comparison, only 11\% of the type I X-ray sources which are
presenting broad emission lines (empty squares) are lying below
$z=0.5$, which is significantly less than for the type 2. This
result is partially due to optical selection effects, as we did
not identify many type 2 AGN at high redshifts ($z\geq1$).
However, it also means that unabsorbed X-ray bright AGN are less
often found as the redshift $z$ decreases.

\begin{figure}
  \includegraphics[angle=0,width=8cm]{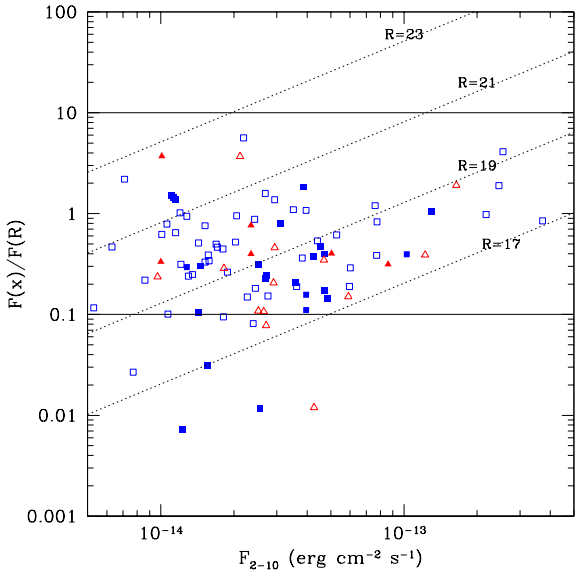}
  \caption{X-ray to optical ratio (2-10 keV band vs $R$ band) as a function of X-ray flux
  for the 94 X-ray sources in our sample for which an $R$ band magnitude is available. Diagonal
  dotted lines indicate loci of constant $R$ magnitude while horizontal solid lines mark the boundary
  lines of the region of canonical AGN. The symbol convention is the same as for Fig.
\ref{lumint}.}
  \label{fluxratio}
\end{figure}

For sources with $L_{2-10}<10^{42}$ erg s$^{-1}$, the presence of
an AGN is no longer certain. Actually these lower luminosity X-ray
sources could be powered by very active star formation galaxies
instead. In these objects, the X-rays may be partially due to the
emission of X-ray binaries, supernovae, and/or to optically thin
thermal emission from hot interstellar gas within the host galaxy.
There are only 3 type I/type 2 sources (namely sources \#13, \#55
and \#63) with X-ray luminosity below $10^{42}$ erg s$^{-1}$. One
of them (source \#63) is embedded in a galaxy cluster detected in
the X-rays, and so the difference in classification for the X-ray
and the optical might be due to the soft emission from the
cluster, which has a temperature around 0.6 keV (see Pierre et al.
2006). These 3 X-ray sources with $L_{2-10}<10^{42}$ erg s$^{-1}$
have an X-ray/optical flux ratio well below 0.1, as can be seen in
Fig. \ref{fluxratio} (the three filled squares). So a priori, they
are more likely starbursts than AGN (Fiore et al. 2003).

Ranalli et al. (2003) have presented tight linear relations
between the X-ray, radio and infrared luminosities of a
well-defined sample of star-forming galaxies. We have used their
relations in order to test the starburst hypothesis. We thus
checked for radio emission using the radio catalogs from Cohen et
al. (2003) and the one from Bondi et al. (2003), which is lying in
the central 1 deg$^{2}$ area of the VVDS area. For these three
starburst candidates, only one (source \#63) has a radio
counterpart (1 mJy at 1.4 GHz). We infer a radio luminosity
$L_{1.4 GHz}=1.65\times10^{29}$ erg s$^{-1}$ Hz$^{-1}$ from its
redshift $z=0.054$. With an X-ray luminosity of
$L_{2-10}\sim2\times10^{41}$ erg s$^{-1}$, it does not follow the
linear relation from Ranalli et al. (2003) (i.e its X-ray
luminosity is about 8 times higher than inferred from its radio
luminosity, which is much larger than expected for star-forming
galaxies, supporting the assumption that this source is an AGN).
Using Fig. 13 of Polletta et al. (2007), source \#63 is neither on
the star-forming galaxies nor on the radio-quiet correlations. It
will be shown in Sect. 5.1.2 that this X-ray source is in fact a
Seyfert 2, based on emission line ratios in its optical spectrum.

Source \#13 has the lowest X-ray luminosity of our sample
($L$$_{2-10}$$\sim$10$^{40}$ erg s$^{-1}$), along with a very low
X-ray$/$optical flux ratio, typical of star-forming galaxies. Thus
this X-ray source is more likely a starburst.

Source \#55 has an X-ray luminosity $L_{2-10}\sim9\times10^{41}$
erg s$^{-1}$, which is pretty close to the AGN threshold. So it
could be either an AGN, or a star-forming galaxy.

For the remaining 22 objects with $L$$_{2-10}$$>$10$^{42}$ erg
s$^{-1}$, the starburst hypothesis is no longer valid, as the
presence of an AGN is unambiguous. These 22 X-ray sources are thus
likely AGN dominated objects. Furthermore 20$/$20 (only 20/22 have
an $R$ magnitude) have an X-ray/optical flux ratio between 0.1 and
10 (Fig. \ref{fluxratio}), which are typical values for AGN (Fiore
et al. 2003). Note that source \#44 is also embedded in a galaxy
cluster and is actually the cD galaxy of the cluster. For this
source, the mismatched classification between the X-ray and the
optical might also be due to additional soft emission from the
galaxy cluster, its temperature being around 2 keV (Pierre et al.
2006).

Moran et al. (2002) have obtained integrated (i.e. nucleus+host
galaxy) optical spectra of a sample of nearby Seyfert 2 galaxies
absorbed in the X-rays band. They find that, due to the
limitations of optical spectroscopic observations, X-ray absorbed
and optically type 2 AGN could be easily undetected in the optical
band. They suggested that host galaxy dilution is a possible
explanation for the lack of AGN lines in their sample of AGN
optical spectra.

More recently, Severgnini et al. (2003) and Silverman et al.
(2005) found a significant number of unabsorbed X-ray sources that
only present narrow emission lines in their optical spectra. They
argued that it could also be attributed to dilution of the AGN
emission by the host galaxy light. In other words, in these AGN,
the nuclear component will be outshone by the optical light of the
host galaxy.

Page et al. (2006a) tested this hypothesis and found that for more
than half of their obscured AGN in the optical, the nuclear
component is indeed outshone by the host galaxy by factors of
3-10. And so they claim that the lack of broad optical emission
lines could be due to the low contrast of the emission lines
against the much stronger starlight component of the host galaxy.

Concerning our 22 AGN, the dilution hypothesis is possible but not
absolutely certain. As it can be seen in Fig. \ref{fluxratio},
there is a small trend for these objects (filled squares) to have
both a brighter optical magnitude and a lower X-ray$/$optical flux
ratio than the unabsorbed sources which are type 1 AGN (open
squares), thus supporting the dilution hypothesis. However this
trend is not very significant: a K-S test has shown that the
X-ray$/$optical flux ratio of these 22 AGN is consistent with the
one which are type 1 AGN ($P\sim0.4$), and that their optical $R$
magnitude is barely significantly brighter than for the type 1 AGN
($P\sim0.07$). Therefore, this diagram alone, is not enough to
argue in favor of the dilution hypothesis to account for their
classification mismatch.

\begin{figure}
  \includegraphics[angle=0,width=8cm]{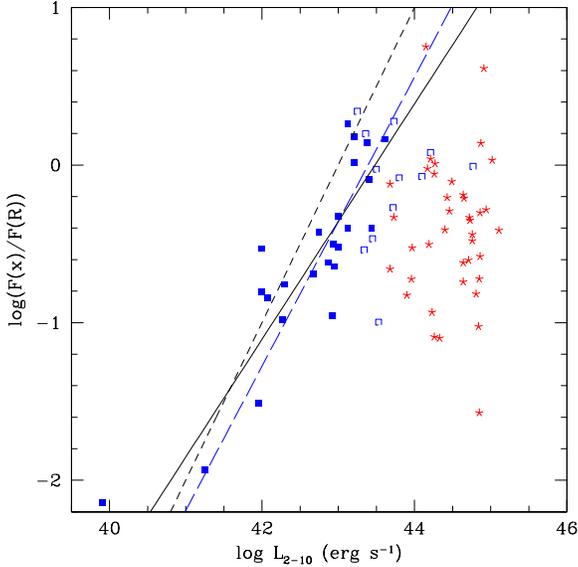}
  \caption{X-ray to optical ratio (2-10 keV band vs $R$ band) as a function of the
  deabsorbed, intrinsic 2-10 keV luminosity for the type I objects.
  Some of them are only presenting narrow emission lines (filled squares). The
  type I that show broad and obvious emission lines in their optical
  spectra are divided into two categories : those which lie at $z<0.8$ (empty squares)
  and those at $z>0.8$ (stars). The solid line represents the
  minimum $\chi^{2}$ fit between log(F$_{\textrm{x}}/F_{\textrm{opt}}$) and log $L_{2-10}$ for
  our data. The short dashed line represents the best linear
  regression for type 2 AGN, obtained using the data of Fiore et al.
  (2003), and the long dashed line is from the data of Treister et al. (2005).}
  \label{fluxratiolum}
\end{figure}

In order to test dilution effects in further details, we computed
the X-ray to optical ratio (2-10 keV band vs $R$ band) as a
function of the unabsorbed, intrinsic 2-10 keV luminosity, as
suggested by Fiore et al. (2003) and Eckart et al. (2006). This
relation is investigated in Fig. \ref{fluxratiolum}, which is
restricted to sources with $z<0.8$ because at $z>0.8$, the $R$
band no longer covers the main part of the stellar component
longward of the Balmer break (empty+filled squares). However, we
also show the unabsorbed type 1 X-ray sources lying at $z>0.8$
(stars) as a matter of comparison. We performed a linear fit to
those objects that lack broad emission lines in their optical
spectra (filled squares) having $L_{2-10}>10^{40}$ erg s$^{-1}$.
We obtain a correlation with ($\chi^{2}_{\nu}=0.09$) using the
$\chi^{2}$ test. We have then compared this value to the
$\chi^{2}_{\nu}$ value obtained if the data are fitted by a
constant ($\chi^{2}_{\nu}=0.29$). We then performed an F-test and
this test showed that the linear fit is statistically a better
description than a constant at a highly significant level
($P\sim5.7\times10^{-7}$). The best-fit parameters obtained for
the linear fit are given by

\begin{equation}\label{chifit}
\log L_{\textrm{x}}=0.747(\pm0.105)\times
\log(\frac{f_{\textrm{x}}}{f_{\textrm{opt}}})-32.48(\pm4.48),
\end{equation}

where $L_{\textrm{x}}$ is the intrinsic X-ray luminosity expressed
in erg s$^{-1}$ and $\frac{f_{\textrm{x}}}{f_{\textrm{opt}}}$ is
the X-ray to optical flux ratio. This correlation is shown by the
solid line in Fig. \ref{fluxratiolum}. The objects that lack broad
emission lines in their optical spectra (filled squares) thus show
a correlation between
log($\frac{f_{\textrm{x}}}{f_{\textrm{opt}}}$) and
log($L$$_{2-10}$), which is expected if the optical light comes
predominantly from the host galaxy as opposed to emission from the
AGN. Fiore et al. (2003), who found a similar correlation, argue
that the observed correlation for the type 2 AGN indicates that
the optical light is largely dominated by the host galaxy, due to
obscuration. A similar correlation has also been reported by
Treister et al. (2005). Indeed, as the X-ray luminosity is mainly
coming from the AGN, for low luminosity AGN, the more luminous the
AGN is in the X-rays, the larger the ratio
$\frac{f_{\textrm{x}}}{f_{\textrm{opt}}}$. Note that this observed
correlation could be either due to obscuration within the AGN, or
to dilution by the host galaxy light. However, as it will be shown
in Sect. (5.1.2), dilution effects are more likely than
obscuration, as most of the optical spectra are representative of
the host galaxy. However, when the AGN is getting much more
luminous, its optical light is not diluted anymore by the host
galaxy. So for these luminous AGN both the optical and the X-ray
components are mainly produced by the AGN. Thus the ratio
$\frac{f_{\textrm{x}}}{f_{\textrm{opt}}}$ is remaining
approximately constant as the X-ray luminosity of the AGN is
increasing.

\begin{figure*}[ht!]
\sidecaption
\includegraphics[angle=0,width=6.5cm]{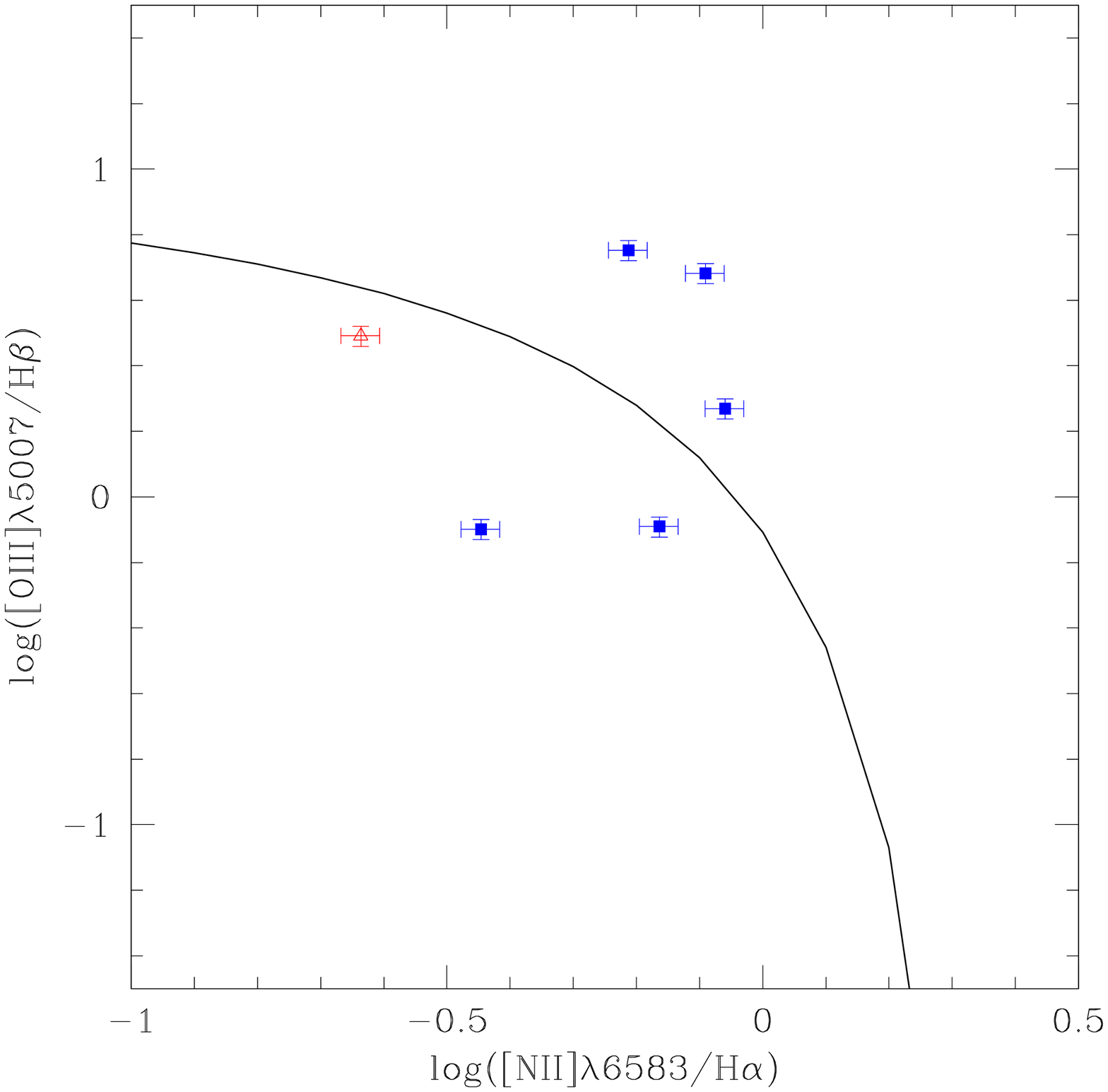}
\includegraphics[angle=0,width=6.5cm]{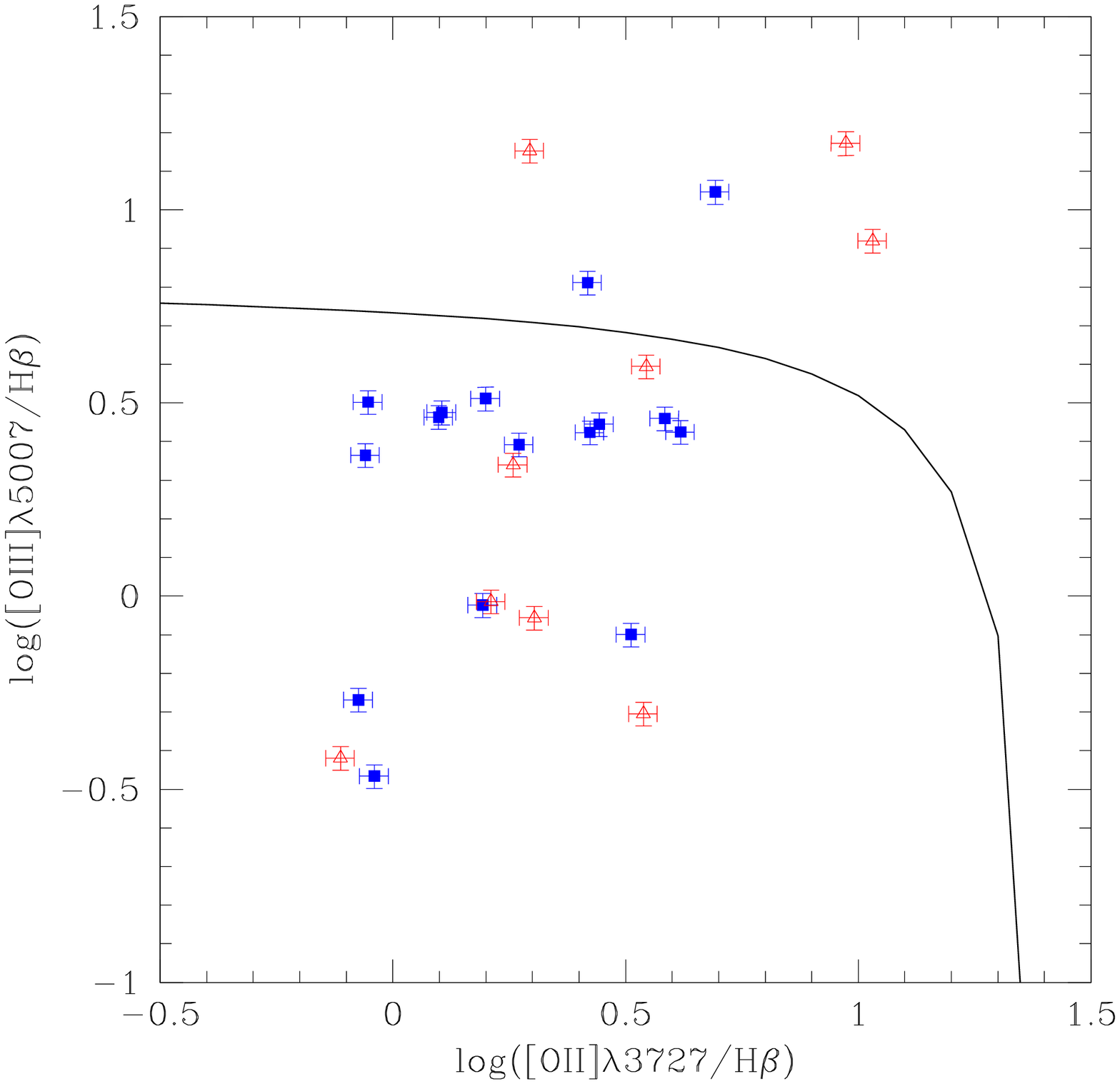}
\caption{Diagnostic diagrams based on line ratios for our
subsample of narrow line AGN with no absorption in the X-rays
(filled squares) and with absorption in the X-rays (empty
triangles). 6 of them for which the H\textrm{$\alpha$} emission
line has been observed, are plotted on the left panel and the
other ones are plotted on the right panel. The solid lines from
Kewley et al. (2006) (left) and from Lamareille et al. (2004)
(right) indicate the separation between star forming galaxies and
AGN, the star forming galaxies being below the line and AGN
(Seyfert+LINER) above. The error bars have been estimated assuming
a 5\% uncertainty for the derived intensities.} \label{lineratio}
\end{figure*}

This is the reason why the X-ray unabsorbed type 1 do not show the
correlation and are clustered at an X-ray luminosity which is
significantly higher than the one for the X-ray unabsorbed  type 2
AGN. We performed a K-S test in order to test the possible
difference in X-ray luminosity for X-ray unabsorbed sources which
are either of type 1 or of type 2. The two sub samples are highly
significantly different, the probability that the two are drawn
from the same population being almost zero. The type 1 objects
have a mean X-ray luminosity log($L$$_{2-10}$)=44.28$\pm$0.55 and
a mean redshift $z$=1.38$\pm$0.66, while the type 2 have a mean
X-ray luminosity log($L$$_{2-10}$)=42.68$\pm$0.83 and a mean
redshift $z$=0.32$\pm$0.23. The unabsorbed X-ray sources which are
classified as type 2 objects are thus both significantly less
luminous than the type 1 objects and are lying at a significantly
smaller redshift. Therefore the dilution hypothesis might not be
ruled out and is, at least statistically, capable to account for
the difference between the X-ray and the optical classifications
for these types of X-ray sources. Therefore, the dilution of the
AGN light by the host galaxy is mainly occurring at small
redshift, supporting the fact that the intrinsic luminosity of the
AGN has significantly declined as the redshift is decreasing.\\
Page et al. (2003) suggested that if the dilution hypothesis holds
true, it may explain why a large fraction of unabsorbed X-ray
sources lacking broad emission lines in their optical spectra are
lying at a low redshift (the majority of our sources of this kind
are lying at $z<0.5$). The characteristic luminosity of AGN has
declined dramatically (by a factor 10) since $z=2$ in both the
X-ray (Page et al. 1997) and in the optical (Boyle et al. 2000).
So unless the host galaxies of AGN have declined in luminosity by
a similar amount, the contrast between the AGN and the host galaxy
light will become smaller and smaller as the redshift decreases.
The trend that more luminous AGN peak at an earlier era, while the
less luminous ones arise later, known as cosmic down-sizing, has
also been reported by other authors (see e.g. Eckart et al. 2006;
Akylas et al. 2006).

\subsubsection{Line ratio diagrams}

In order to better understand the nature of the unabsorbed X-ray
sources, which are only showing narrow emission lines in their
optical spectra (type I X-ray sources which are type 2), we have
used a refined classification based on the optical spectra. This
is based on the diagnostic diagrams of Lamareille et al. (2004),
who use blue emission lines ([OII] $\lambda$3727, [OIII]
$\lambda$5007 and H$\beta$) to discriminate star forming galaxies
and HII regions from AGN among intermediate-redshift ($z>0.3$)
objects. This constitutes an improvement compared to the diagram
of Veilleux \& Osterbrock (1987) who only used
[OIII]$\lambda$5007$/$H$\beta$ vs [NII]$\lambda$6583$/$H$\alpha$,
since these two line ratios are only efficient at very low
redshifts ($z<0.2$). Note that very recently, Kewley et al. (2006)
have obtained a refined optical classification between pure star
forming galaxies, composite galaxies (star forming region+AGN),
LINERs and Seyferts. Their refined classification is based on
three diagnostic diagrams, namely [OIII]$\lambda$5007$/$H$\beta$
vs [NII]$\lambda$6583$/$H$\alpha$; [OIII]$\lambda$5007$/$H$\beta$
vs [SII]$\lambda$6717$/$H$\alpha$ and
[OIII]$\lambda$5007$/$H$\beta$ vs [OI]$\lambda$6300$/$H$\alpha$.
However, these diagrams are only efficient at very low redshifts,
as the initial diagram of Veilleux \& Osterbrock (1987). We ran
the task \texttt{splot} in the \textsc{IRAF} package
\texttt{onedspec} to compute the line ratios. We used the diagrams
of both Kewley et al. (2006) and Lamareille et al. (2004),
whenever possible. Most of the time the H$\alpha$ emission line
was not present because of the spectral coverage and thus we
mainly used the line ratios suggested by Lamareille et al. (2004):
[OIII] $\lambda$5007$/$H$\beta$ vs [OII] $\lambda$3727$/$H$\beta$.

Fig. \ref{lineratio} gathers the line ratios for narrow line AGN
presenting both unabsorbed X-ray spectra (filled squares) and
absorbed X-ray spectra (empty triangles). More specifically, among
the 22 X-ray sources upon which we test the dilution hypothesis
(22 X-ray sources out of 25 have $L$$_{2-10}$$>$10$^{42}$ erg
s$^{-1}$), 15 (68\%) are pure star forming galaxies, 4 (18\%) are
Seyfert 2, 1 is an absorption line galaxy (ALG, 5\%) and 2 (9\%)
are undefined (because of the lack of sufficient spectral features
to compute line ratios). We have therefore shown that most of the
optical spectra for these kinds of sources are more representative
of stellar processes in the host galaxy rather than due to the
AGN, which constitutes one more argument in favor of the dilution
hypothesis. However, we have also found 4 Seyfert 2 which are
unabsorbed in the X-rays. Unabsorbed Seyfert 2 have already been
found in other samples as the one of Panessa \& Bassani (2002).
They claim that these unabsorbed Seyfert 2 might be weak AGN in
which the broad line region (BLR) is non-standard (very weak or
fading away) or in which the BLR does not exist at all. This
scenario is more likely applicable to low luminosity AGN where the
brightness of the active nucleus may be insufficient to
photoionize the BLR. These 4 Seyfert 2 (i.e \#9, \#53, \#56 and
\#82) have $L$$_{2-10}$$\sim$10$^{43}$ erg s$^{-1}$ and therefore
are not considered as low luminosity AGN.

Next, we also computed the line ratios to investigate in further
details the nature of the 3 starburst candidates, discussed at the
beginning of Section 5.1. We have found that source \#13 is a pure
star forming galaxy. Its optical spectrum is dominated by the
photospheric emission of very hot stars. Thus its X-ray emission
could likely be due to the combination of hot gas from supernovae
and high mass X-ray binaries in a star-forming galaxy. Source \#55
has been classified as an absorption line galaxy, which is not
consistent with the starburst hypothesis. And thus, this X-ray
source is more likely a LLAGN which is heavily diluted by the
light of its host galaxy. Finally, source \#63 has been classified
as a Seyfert 2 from its optical line ratios, instead of a
starburst. This X-ray source is thus a low luminosity AGN similar
to those detected by Panessa \& Bassani (2002), in which the BLR
might be very weak or fading away. On the other hand, the source
\#63 could also be a good Compton thick candidate, as a Fe
emission line has been detected in its X-ray spectrum. In this
case, we only observe the indirect reflected component of the
X-ray emission, and thus both its intrinsic X-ray luminosity and
hydrogen column density would have been highly underestimated. One
way to test the Compton thick nature of this AGN would be to plot
the equivalent width of the Fe line against the
$\frac{F_{\textrm{x}}}{F_{\textrm{[OIII]}}}$ ratio (See Fig. 1 of
Bassani et al. 1999). Unfortunately, we have not been able to
compute $F_{\textrm{[OIII]}}$, as our optical spectra are only
calibrated in relative fluxes. Thus, out of the 3 starburst
candidates, only 1 is a starburst galaxy. Of course, the above 4
unabsorbed Seyfert 2 could also be Compton thick candidates.
However, as no Fe line have been detected in their X-ray spectra,
this hypothesis is much less probable to take into account their
nature.

\begin{table}
\begin{tabular}{|c|c|c|c|}
\hline
   & Type I 2 & Type II 2 & N\\
  \hline
  \textbf{SF glx} & 16 & 5 & 21 \\
  \textbf{Sy 2} & 5 & 5 & 10 \\
  \textbf{ALG} & 2 & 1 & 3 \\
  \textbf{Undef} & 2 & 2 & 4 \\ \hline
    N & 25  &  13 & \textbf{38} \\
    \hline
\end{tabular}
\caption{Refined classification of the type 2 sources (both type I
and type II), which is based on the emission line ratios computed
from their optical spectra.} \label{tableline}
\end{table}

\begin{figure*}[ht!]
\includegraphics[angle=270,width=7.5cm]{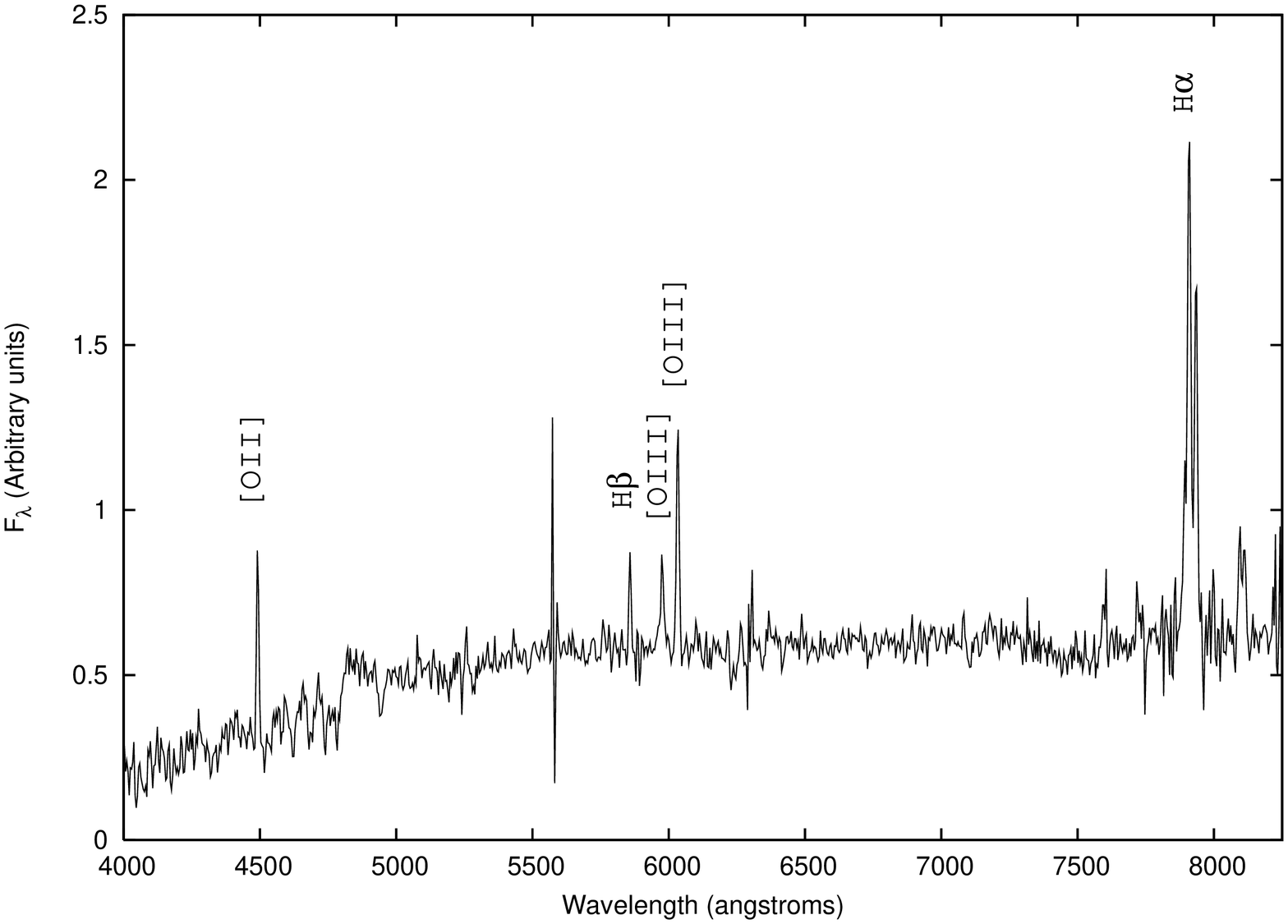}
\includegraphics[angle=270,width=7.5cm]{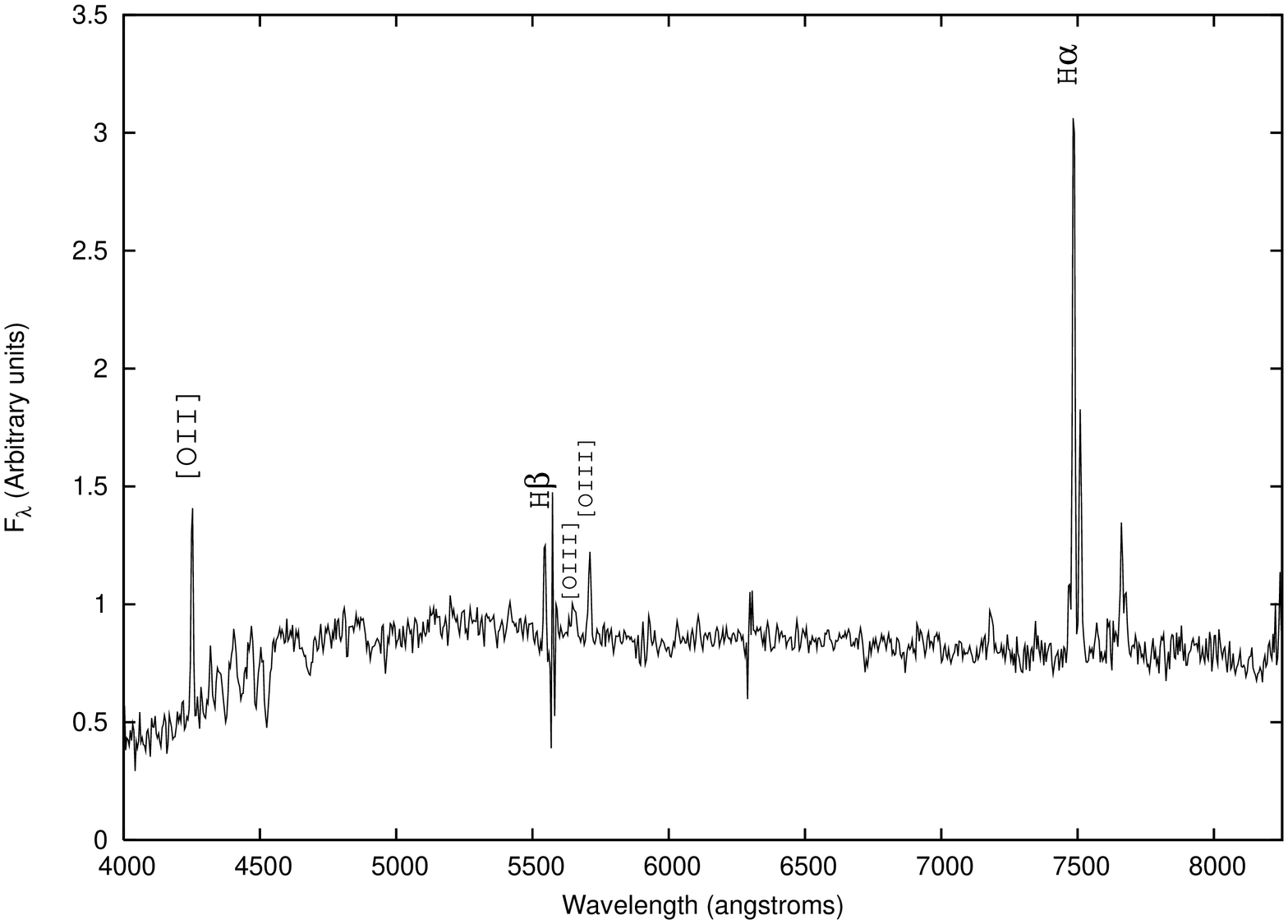}
\includegraphics[angle=270,width=9cm]{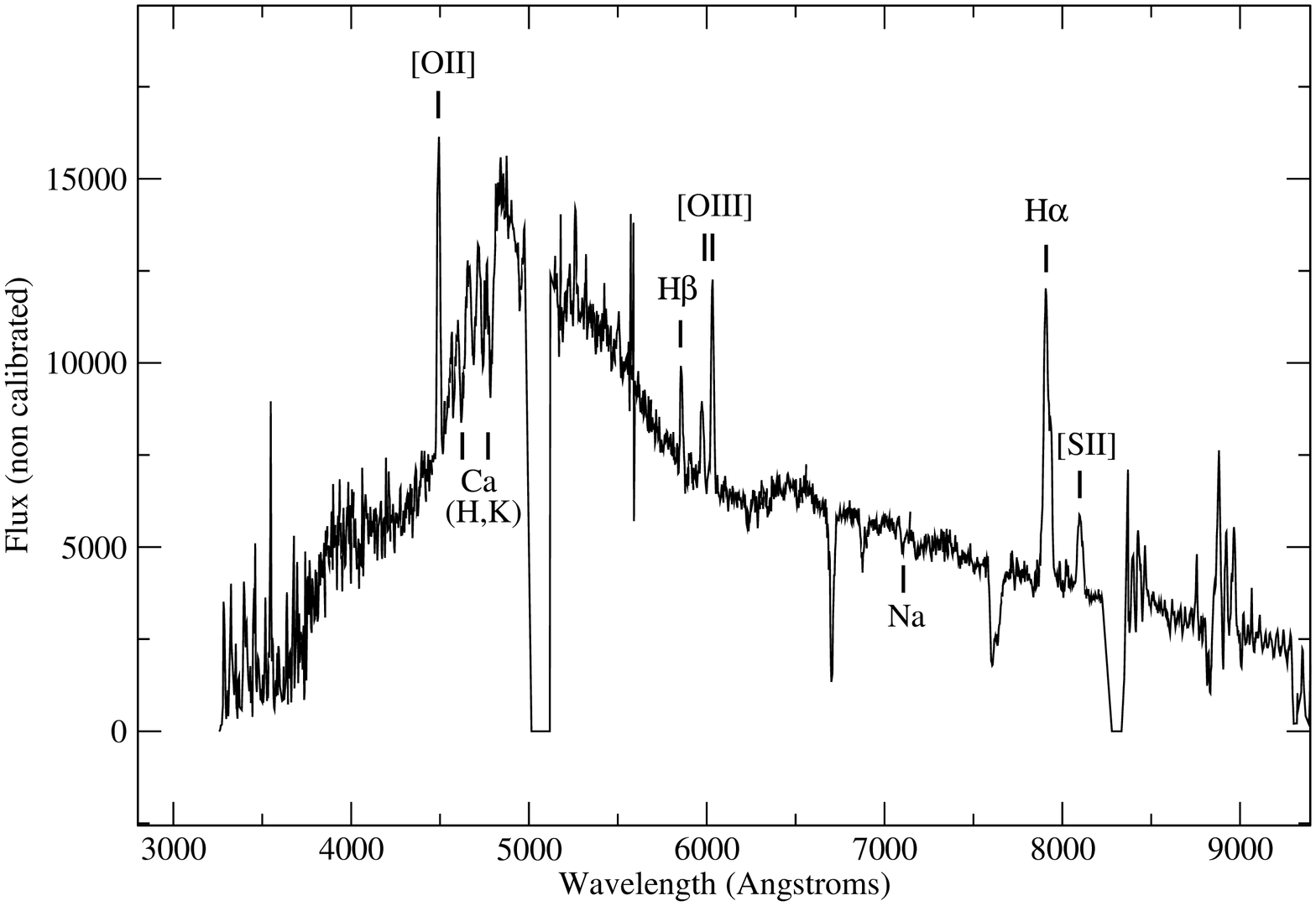}
\hspace{1.7cm}
\includegraphics[angle=270,width=9.1cm]{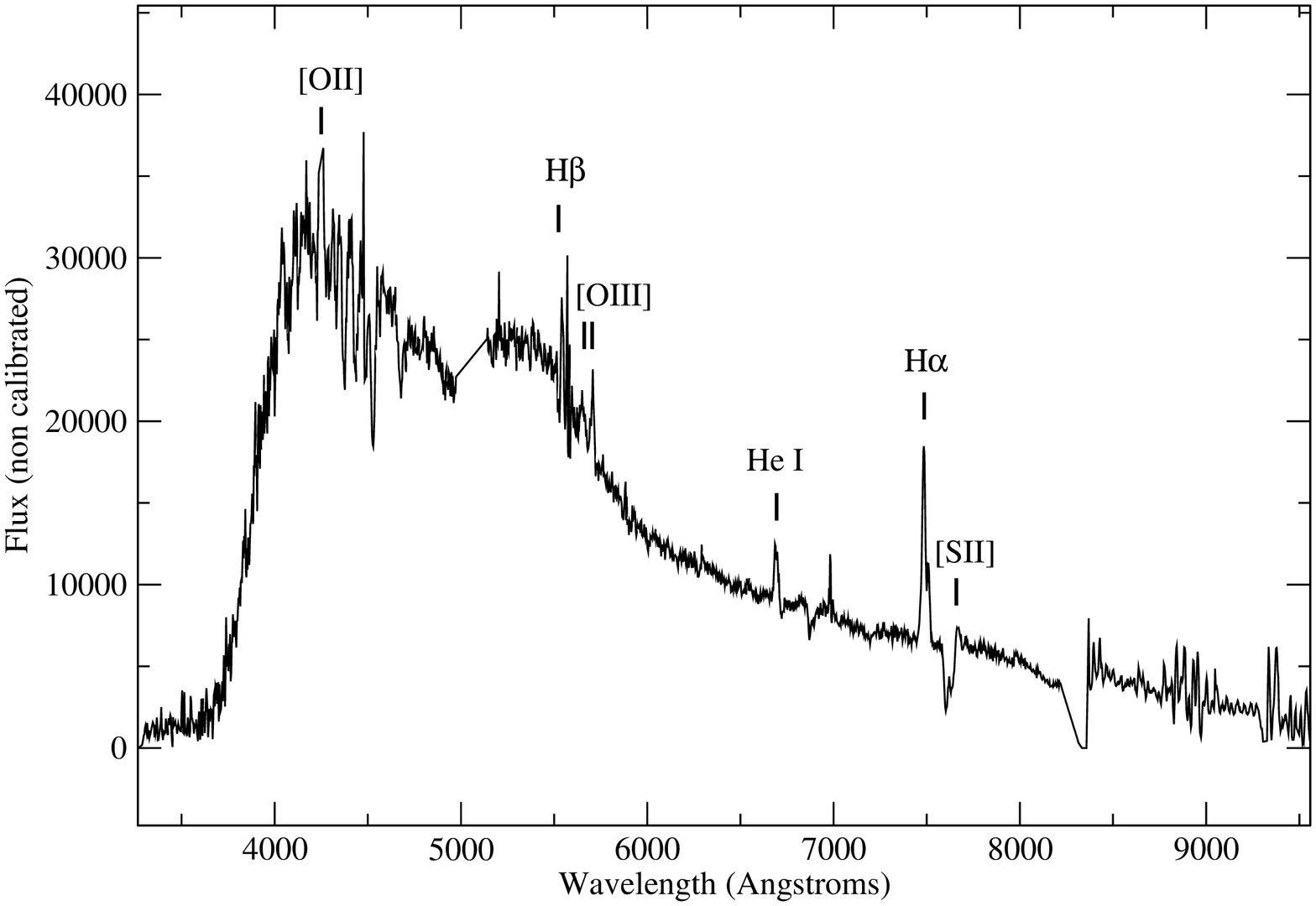}
\caption{2dF spectra of two absorbed X-ray sources which only
present narrow emission lines in their optical spectra (upper
panel). These are sources \#84 and \#66 in Table 4, respectively.
They have been classified as star forming galaxies, based on the
line ratios inferred from their optical spectra. The lower panel
show the optical spectra of the corresponding objects, obtained
with the SALT. Both have been observed with the PG0300 grism
($R\sim400$). The SALT optical spectra are not flux calibrated.}
\label{class22}
\end{figure*}

Finally, we have computed the line ratios for the absorbed X-ray
sources which are only presenting narrow emission lines in their
optical spectra (type II X-ray sources which are type 2). As can
be seen in Table 2, there are 13 such sources. 5 of them (38\%)
are Seyfert 2. Fig. \ref{lineratio} shows only 3/5 in the AGN
region of the diagrams. However, there are also two (empty
triangles) slightly below the dividing line (thus in the star
forming galaxy region) that have been classified as Seyfert 2
anyway, because of the presence in their optical spectra of the
emission line of [NeIII]$\lambda$ 3869, which is a line typical of
AGN. 1 is an absorption line galaxy (8\%), 5 are pure star forming
galaxies (38\%), and 2 are unidentified (16\%). Note that 1 of the
2 unidentified objects is a NELG, and coupled with its X-ray
characteristics has been classified as a type II QSO. So, for a
majority of these 13 X-ray sources (7/13), the optical spectra are
AGN-like, rather than resembling the one of the host galaxy.
Therefore the absorbed X-ray sources which are only presenting
narrow emission lines are truly hidden AGN.

Table 3 shows an overview of the refined classification of the
obscured X-ray sources, which is based on the emission line
ratios.

Finally, Fig. \ref{class22} shows two examples of absorbed X-ray
sources which only present narrow emission lines in their optical
spectra.

\subsubsection{Testing the dilution hypothesis}

Until now we have presented several observational arguments in
favor of the dilution of the AGN optical light by the host galaxy
light. We now develop a simple model in order to investigate
whether dilution effects are plausible or not. In this model, we
consider the contributing optical flux in a 1 $\arcsec$ slit from
the host galaxy and, first, from X-ray unabsorbed AGN only
presenting narrow emission lines ($L_{2-10}\sim5\times10^{42}$ erg
s$^{-1}$; $z\sim0.3$) and from X-ray unabsorbed AGN presenting
broad emission lines ($L_{2-10}\sim2\times10^{44}$ erg s$^{-1}$;
$z\sim1.5$). In order to convert X-ray flux into optical flux, we
take a typical ratio $\frac{f_{\textrm{x}}}{f_{\textrm{opt}}}=1$.
Next, we model the host galaxy with $M_{R}=-22.$ ($R\sim19$ for
$z\sim0.3$) and with a physical size of 30 kpc. Finally, we apply
the corresponding k-correction for a typical elliptical galaxy
(Coleman et al. 1980), and we use the galaxy profile of Burkert et
al. (1993) in order to estimate the fraction of the total optical
flux of the host galaxy which enters the slit. Making the
reasonable assumption that the typical absolute magnitude of the
host galaxy of radio quiet AGN is similar at $z=0.3$ and at
$z=1.5$, (Kukula et al. 2001, Fig. 7), we derive the following
values:

$-$ At $z=0.3$, \begin{equation}\label{host}
(\frac{F_{\textrm{host,slit}}}{F_{\textrm{AGN}}})_{R}\sim2,
\end{equation}

$-$ At $z=1.5$, \begin{equation}\label{hostbis}
(\frac{F_{\textrm{host,slit}}}{F_{\textrm{AGN}}})_{R}\sim\frac{1}{400},
\end{equation}

where $F_{\textrm{host,slit}}$ is the fraction of the host galaxy
light entering the slit and $F_{\textrm{AGN}}$ is the AGN light
entering the slit. Note that other previous works have claimed
that the host galaxies of AGN were about 2.5 mag brighter at
$z\simeq2$ than those of low-redshift AGN (see e.g. Aretxaga et
al. 1998). But even taking this into account, the AGN at
$z\simeq1.5$ would still be about 40 times brighter than their
host galaxies, thus preventing dilution effects.

Thus for the above AGN around $z=0.3$, the flux of the host galaxy
entering into the slit is about twice as large as the optical flux
of the AGN. On the other hand, for high redshift AGN ($z=1.5$),
the optical flux of the AGN is totally overwhelming the optical
flux of the host galaxy and dilution effects can not occur
anymore. Even if crude, this model allows us to confirm that
dilution can definitely be present in low luminosity and low
redshift AGN, if host galaxy properties do not evolve
significantly with the redshift.

\begin{figure*}[ht!]
\centering
\includegraphics*[angle=270,width=8.8cm,bb=573 43 77
  721]{7778fig15a.ps}
  \includegraphics*[angle=0,width=9.5cm]{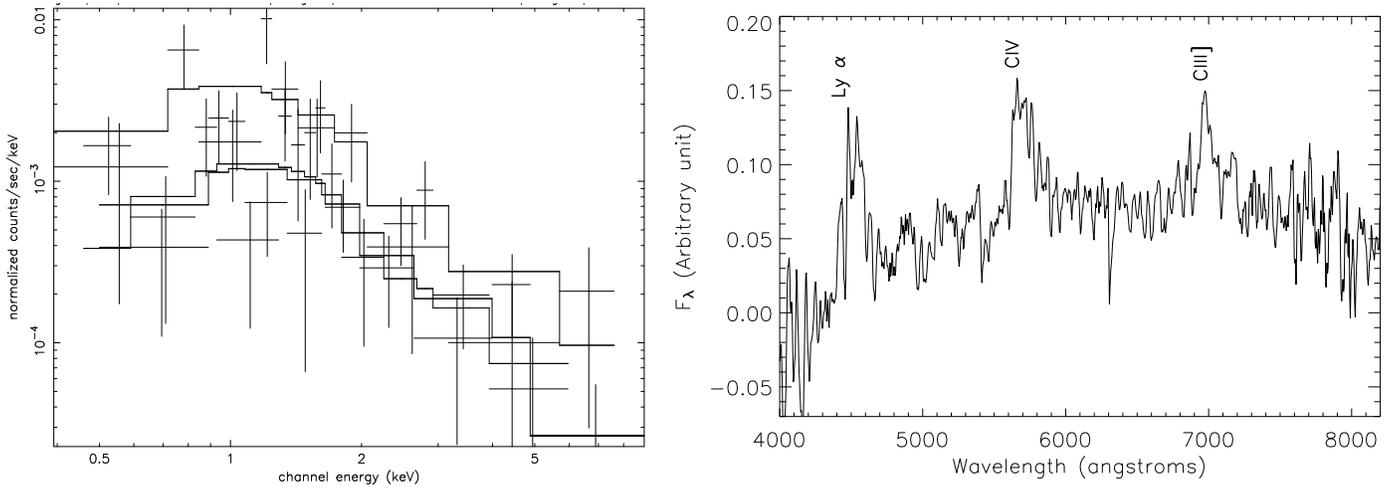}
  \caption{MOS1 MOS2 (the two lower functions) and pn (upper one) X-ray spectra of a source which is showing strong X-ray absorption
  with $N_{\mathrm{H}}^{\mathrm{int}}$$=$4 10$^{22}$ cm$^{-2}$. The crosses represent the data points and the solid
  line represent the folded model simultaneously fitted to the data of the
  3 detectors (left). The optical counterpart has been identified as a
  BLAGN lying at a redshift $z=2.66$ (right). This is source \#20 in Table 4.}  \label{class21}
\end{figure*}

As a conclusion, in this section, we have shown that dilution
effects can very possibly explain the difference between the X-ray
and the optical classifications of a great majority of these 25
X-ray sources. Among these 25 X-ray sources, we have also found 5
Seyfert 2 (4 with $L$$_{2-10}$$>$10$^{42}$ erg s$^{-1}$, and 1
with $L$$_{2-10}$$<$10$^{42}$ erg s$^{-1}$)  which do not present
strong X-ray absorption in their X-ray spectra.

Therefore, apart from these 5 X-ray unabsorbed Seyfert 2, the
dilution hypothesis works, without any need to alter the standard
orientation-based AGN unified scheme.

\subsection{Absorbed AGN showing broad emission lines in their
optical spectra}

Fig. \ref{class21} illustrates a representative example of an
absorbed X-ray source with bright and obvious broad emission lines
in its optical spectrum and so with no hint of obscuration in the
optical. This is source \#20 in Table 4. We have found 7 such
X-ray sources.

These 7 X-ray sources are only found in the high luminosity regime
of AGN ($L$$_{2-10}$$>$6$\times$10$^{43}$ erg s$^{-1}$), 4 of them
having an $L$$_{\mathrm{X}}$ typical of QSOs. Because of their
high X-ray luminosity, dilution effects are totally ruled out to
account for their nature. Therefore these 7 objects are not
consistent with the standard orientation-based unified scheme.
They span a broad range in redshift ($z$=[0.4-3.6]), most of them
being high redshift AGN.

In Fig. \ref{fluxratio}, these 7 sources (filled triangles) have
X-ray$/$optical ratios typical of AGN which are both unabsorbed in
the X-rays and which present broad emission lines in their optical
spectra (empty squares). In order to investigate this in further
details, we have compared the average X-ray luminosity of these
absorbed AGN being type 1 with the one of the unabsorbed AGN, also
being type 1. The average X-ray luminosities are log
$L$$_{2-10}$=44.59$\pm$0.56 and log $L$$_{2-10}$=44.28$\pm$0.55,
respectively, which turns out to be quite similar. We also
compared the average spectroscopic redshifts for these 2 classes.
We obtain $z=1.93$ and $z=1.38$, respectively, both representative
of high redshift AGN. The above 2 classes of objects have on
average the same intrinsic X-ray luminosity and redshift, and are
both lying in the high luminosity/high redshift regime. These two
types of objects thus seem to belong to the same AGN population.

Such anomalous objects have already been observed by several
authors (e.g Perola et al. 2004; Eckart et al. 2006). They
respectively find that around 5\% and 15\% of the X-ray sources in
their respective sample are X-ray absorbed sources which present
broad emission lines in their optical spectra. Both these values
bracket the fraction that we have found, which is around 7\%.
Interestingly, these two studies also find that these X-ray
absorbed sources with broad emission lines in their optical
spectra are high luminosity AGN lying at high redshift (typically
$z>1$). So it seems that high luminosity and high redshift are
common properties of these peculiar objects.

Several interpretations have been suggested in order to explain
their nature. Maiolino et al. (2001a) have presented a sample of
19 nearby AGN whose X-ray spectra show evidence for cold
absorption and no hint of obscuration in the optical (hence
classified as type 1). They concluded that the
$\frac{E_{B-V}}{N_{\textrm{H}}}$ ratio is systematically much
lower than the Galactic standard value. In a companion paper,
Maiolino et al. (2001b) suggest that a dust distribution dominated
by large grains in the obscuring torus could explain the low
$\frac{E_{B-V}}{N_{\textrm{H}}}$ values obtained. They claim that
the formation of large grains is naturally expected in the high
density environment characterizing the circumnuclear region of
AGN. These large grains make the extinction curve flatter than the
galactic one and thus for a given $N_{\textrm{H}}$ a reduced
extinction and reddening are observed, compared to the galactic
standard.

Alternatively, Weingartner $\&$ Murray (2002) propose a model to
explain these absorbed AGN presenting broad and obvious emission
lines in their optical spectra. They show that the large grain
hypothesis in the obscuring torus is not needed in order to
explain most of the sample of Maiolino et al. (2001a): they
suggest that the material that absorbs the X-rays is probably
unrelated to the material that absorbs the optical/infrared
radiation and that the torus is probably not probed by the
observations of Maiolino et al. (2001a). They suggest that the
line of sight of the sample of Maiolino et al. (2001a) passes
through ionized material located just off the torus and/or
accretion disk. This material is responsible for the X-ray
absorption, while the optical/infrared extinction occurs in
material farther from the nucleus, where the dust may be quite
similar to galactic dust. The X-ray-absorbing material may be
dust-free or may contain large grains that have very small
extinction efficiencies in the optical/infrared. This material may
be associated with a disk wind, which would originate within the
dust sublimation radius (see Murray et al. 1995). In this case,
the dust will sublimate and the obscuration/extinction in the
optical will be much reduced, even if there is a strong absorption
in the X-rays, produced by the ionized gas. Note that according to
this model, the optical spectra of this type of sources should
also present broad absorption lines, and thus be classified as BAL
QSOs, because of the wind outflow.

Recent works suggest that BAL QSOs (which are high luminosity AGN
with broad absorption and emission lines in their optical spectra)
are weak soft X-ray sources, the X-ray weakness being attributed
to absorption rather than being intrinsic (Brandt et al. 2000;
Punsly 2006). Thus most of the BAL QSOs can also be classified as
X-ray absorbed sources which present broad emission lines in their
optical spectra. Therefore, it is tempting to suggest that these
QSOs (which are optically selected) are actually the same
population of objects than the X-ray absorbed type 1 AGN, detected
in X-ray surveys.

However, generally, only up to a few percent of the X-ray absorbed
sources with broad emission lines are identified as a BAL QSO:
none of our 7 X-ray sources have an optical spectrum consistent
with a BAL QSO spectrum. Similarly, Perola et al. (2004) do not
report any BAL QSOs among the X-ray absorbed type 1 AGN from their
sample. Finally, Eckart el al. (2006) find that only 14\% of these
sources are BAL QSOs. Thus generally, these two types of AGN do
not seem to belong to the same AGN populations. Indeed, Risaliti
et al. (2001) have shown that only a fraction of the absorbed
X-ray sources with broad emission lines, detected in X-ray
surveys, could also actually be classified as BAL QSOs, with the
latter constituting the tail of the population of X-ray weak
quasars. However, note that, as BAL occur mainly in the UV,
blueward of the CIV emission line, the spectral coverage of the
optical spectra does only allow the observation of the broad CIV
absorption line for high z AGN (typically $z>1.6$ for the blue
part of the spectrum starting at 4000 \AA). Moreover, the broad
absorption lines could be difficult to detect due to the moderate
level of extinction and generally require high signal-to-noise
ratio optical spectra. Thus the fact that X-ray absorbed type 1
AGN and BAL QSOs have common properties is not totally ruled out.

One way to choose between the large grain hypothesis and the low
dust-to-gas ratio would be to fit the X-ray spectra of the 7
objects with models including gas being highly ionized and check
if they fit better the X-ray spectra than a model with neutral
gas. Unfortunately, we have not been able to test it, as our X-ray
spectra do not have a signal to noise ratio high enough.

Alternatively, these 7 X-ray sources could be at a different
evolutionary stage : the fact that the absorbed AGN presenting
broad emission lines in their optical spectra are only found in
the high luminosity/redshift regime of the AGN is consistent with
the scenario suggested by Page et al. (2004). They have observed 8
absorbed AGN presenting broad emission lines in their optical
spectra in the submillimeter domain with SCUBA and showed that
these QSOs are characterized by a much stronger submillimeter
luminosity compared to a sample of 20 unabsorbed AGN presenting
broad emission lines in their optical spectra. They claim that
this can be understood as an isotropic signature of copious star
formation, with the QSO being embedded in the dense interstellar
media of their forming host spheroids. This rules out orientation
effects as the cause of the absorption. Very recently, Page et al.
(2006b) have suggested that the X-ray absorption of these objects
is most likely due to a dense ionized wind driven by the QSO. This
wind could be the mechanism by which the QSO terminates the star
formation in the host galaxy, and ends the supply of accretion
material, to produce the present day black hole/spheroid mass
ratio.

Therefore they suggest that the absorbed and unabsorbed QSOs
represent different stages in an evolutionary sequence, in which
the absorbed QSOs represent the earlier phase. They estimate that
the absorbed phase typically lasts only around 15\% of the
unabsorbed phase. In our sample, we have found 61 type 1 AGN. 7 of
them are presenting strong X-ray absorption. So these objects
represent about 11\% of the type 1 AGN. This is consistent (within
1 $\sigma$) with the value predicted by Page et al. (2004).
However, we would need to detect many more objects of this kind in
order to better understand their nature and to test in further
details the above model.

We finally compared the X-ray luminosity and redshift for the
X-ray absorbed AGN classified as type 1 (with broad emission lines
in their optical spectra) and type 2 (only with narrow emission
lines in their optical spectra). The mean X-ray luminosities are
log $L$$_{2-10}$=44.59$\pm$0.56 and log
$L$$_{2-10}$=42.89$\pm$0.69, respectively. Concerning the
redshifts, the corresponding mean values are $z=1.93$ and
$z=0.317$, respectively. So, when comparing these two classes, we
notice that they differ significantly, both in redshift and
luminosity: the absorbed AGN presenting broad emission lines in
their optical spectra are both more luminous in the X-rays and
lying at a significantly higher redshift than the ones only
presenting narrow emission lines in their optical spectra.

As a conclusion, we have confirmed the existence of a population
of AGN which are classified as type II in the X-rays and as type 1
in the optical. This population could actually be X-ray unabsorbed
type 1 X-ray sources, intrinsically (either they are at a
different evolutionary stage, or their dust is sublimated; either
way, we can not rule out the presence of an intrinsic type 1 QSO,
surrounded by a torus).

\section{X-ray bright optically normal galaxies (XBONG)}

The existence of an intriguing population of galaxies with X-ray
properties suggesting the presence of an AGN, but without any
obvious sign of activity in their optical spectra, has been known
for more than 20 years (e.g. Elvis et al. 1981; Griffiths et al.
1995; Severgnini et al. 2003). The nature of this kind of sources
is far from being understood. Different scenarios have been
proposed so far: a) the AGN is a BL Lac object; b) the nucleus
activity is outshone by the stellar continuum of the host galaxy
(as shown by Severgnini et al. 2003); c) the emission lines could
be absorbed by material beyond the NLR. Most of these optically
dull galaxies seem to be absorbed in the X-rays (Mainieri et al.
2002). Comastri et al. (2003) have recently shown that the
distribution of X-ray-to-optical flux ratio of XBONG sources could
be well reproduced assuming that the underlying SED of the
putative AGN is that of an X-ray Compton-thick AGN.

\begin{figure}
  \includegraphics[angle=0,width=9cm]{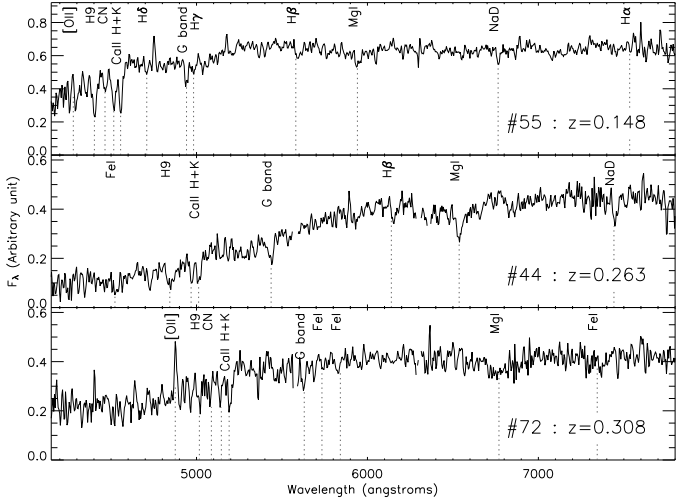}
  \caption{The 3 XBONG candidates. Each of these 3 XBONG lacks emission lines, except
  for [OII] in source \#72. This source presents strong absorption in the X-rays
  ($N_{\mathrm{H}}^{\mathrm{int}}$=3.86$\times$10$^{22}$ cm$^{-2}$).}
  \label{xbong}
\end{figure}

In our sample of X-ray sources, 3/99 have been classified as
absorption line galaxies (See Fig. \ref{xbong}). The X-ray
properties of these 3 X-ray sources suggest the presence of AGN
activity : they have an intrinsic luminosity
$L$$_{2-10}$$>$9$\times$10$^{41}$ erg s$^{-1}$, two of them having
$\frac{F_{\textrm{x}}}{F_{\textrm{opt}}}>0.1$, similar to typical
values of luminous AGN (Fiore et al. 2003). So these 3 X-ray
sources are good XBONG candidates. Two of them (source \#44 and
source \#55) are unabsorbed in the X-rays. Dilution effects could
be responsible for the lack of emission lines in the optical
spectra of these two X-ray sources, as they lie on the relation
given by eq. (4). Finally, source \#72 presents strong absorption
in the X-rays ($N_{\mathrm{H}}^{\mathrm{int}}$$>$ 10$^{22}$
cm$^{-2}$ at the 95\% confidence level). For this source, strong
absorption along with some dilution effects could explain why this
source lacks AGN features in its optical spectrum and is thus
classified as XBONG.

These 3 X-ray sources have similar properties to some XBONG that
are emerging from different X-ray surveys (e.g. Severgnini et al.
2003), although a closer look at their spectra, based on high
resolution data and/or on a better spectral coverage might reveal
their real AGN nature. The finding of these 3 XBONG candidates
clearly shows that optical spectroscopy sometimes can be
inefficient in revealing the presence of an AGN, which instead
becomes obvious from an X-ray spectroscopic investigation.

\section{Type II QSO candidates}

In this work, we have discovered 6 X-ray sources with
$N_{\mathrm{H}}^{\mathrm{int}}$$>$10$^{22}$ cm$^{-2}$ and
$L$$_{2-10}$$>$10$^{44}$ erg s$^{-1}$. However, 5 of these sources
have type 1 optical spectra (e.g. with broad emission lines). The
remaining object (source \#42), classified as a type 2 QSO from
its optical spectrum, is therefore the only genuine type II QSO
detected amongst our sample of 99 X-ray sources. This object is
lying at a rather high redshift ($z\sim1$).

Recently, a connection between Extremely Red Objects (EROs) and
type II QSOs has been suggested (see e.g. Gandhi et al. 2004,
Severgnini et al. 2005). The UKIDSS $K$ band magnitude of our type
2 QSO candidate is $K\sim17.2$ and its color $R\--K$ is equal to
4.6. It is thus very close to the $R-K>5$ threshold adopted to
define the population of EROs. Note that this type II QSO
candidate has independently been reported by Tajer et al. (2007)
in fitting photometric data points with several SED templates. We
confirm here the nature of this source.

\section{Summary and conclusions}

We have presented and characterized a sample of 99 X-ray point
sources for which we have secure identification and spectroscopic
redshift. These 99 X-ray sources have been selected in the [2-10]
keV band with at least 80 counts in the [0.5-10] keV band, from a
sample of 612 X-ray sources.

A large fraction of our X-ray sources (about 95\%) have been
identified as AGN. We have analyzed their optical classification
by measuring the FWHM of the emission lines present in their
optical spectra. In particular we have divided the X-ray sources
in 61 type 1 AGN (BLAGN) and 38 type 2 AGN (35 NELGs+3 ALGs). Next
we have fitted the X-ray spectra of these sources and inferred
their hydrogen column density. Proceeding this way, we have found
79 type I X-ray sources ($N_{\mathrm{H}}^{\mathrm{int}}$ $<$
10$^{22}$ cm$^{-2}$) and 20 type II X-ray sources
($N_{\mathrm{H}}^{\mathrm{int}}$ $\geq$ 10$^{22}$ cm$^{-2}$). The
X-ray and the optical properties have been analyzed and
intercompared. The main results of our work can be summarized as
follows :
\begin{itemize}
  \item We have shown that there is at most a mild correlation
  between the X-ray and the optical classifications, 32 X-ray
  sources out of 99 having different X-ray and optical
  properties. Then we have shown that this discrepancy is not a function
  of the X-ray flux and luminosity. These results agree with
  several other published studies.

  \item We have gathered several pieces of evidence in favor of dilution effects,
  where the AGN light is overshone by a luminous host galaxy, for a vast majority of
  the 25 unabsorbed X-ray sources which only present narrow
  emission lines in their optical spectra. This is a nice confirmation of
  previously published results (e.g Severgnini et al. 2003,
  Silverman et al. 2005). We have also found 5 Seyfert 2 which do not
  present a strong absorption in the X-rays. Only 1 of them is a good
  Compton thick AGN candidate.

  \item We have reported 7 absorbed X-ray sources
  which present broad emission lines in their optical spectra. This
  corresponds to 7\% of the whole sample and to 11\% of the type 1
  AGN. These objects, which are highly luminous AGN (mostly QSOs) and lie at
  large redshifts, have similar X-ray luminosity and redshift
  as the unabsorbed X-ray sources which present broad
  emission lines in their optical spectra. The difference between
  the X-ray and the optical classifications of these 7 X-ray
  sources could easily be accounted for if the gas responsible for the
  X-ray absorption is highly ionized, instead of being neutral, in which case the accompanying dust
  would sublimate to yield a much smaller dust-to-gas ratio, resulting in
  a reduced optical extinction and reddening. However, our X-ray spectra do not have a
  signal to noise ratio high enough to
  distinguish ionized gas model from neutral gas models. Moreover, part of these sources
  could be BAL QSOs. From this perspective, it seems plausible
  that the obscuring material in these sources has different
  composition than typical Seyferts (e.g. either a very low dust:gas
  ratio, or a dust-free wind).

  \item Our study, which has been done in parallel to the work
  of Tajer et al. (2007), has allowed us to confirm their main
  results, but using a safer procedure as we utilized
  spectroscopic redshifts along with secure spectroscopic
  identifications, instead of photometric redshifts obtained by
  fitting photometric data points with SED templates. Moreover, our study covers
  a much larger area.

  \item We have found 1 genuine type II QSO candidate amongst our X-ray sample.
  Note that we have also identified 5 X-ray absorbed QSOs which
  are not obscured in the optical.
\end{itemize}
As a conclusion, most of the discrepancy between the X-ray and the
optical classifications comes from the fact that type 2 optical
sources are more likely to be unabsorbed than absorbed in the
X-rays. We have shown that the vast majority of these type 2
optical sources which are unabsorbed in the X-rays are affected by
dilution effects, which do not require any modification of the
standard orientation-based AGN unification scheme. Thus this AGN
unification scheme still holds for about 88\% of the X-ray sources
in our sample, as their predictions are not met for only 12/99 of
the X-ray sources. Note that the discrepancy between the
absorption properties in the X-rays and in the optical/UV that we
have found, could as well be a natural consequence of the clumpy
nature of the absorbing medium, as suggested by Elitzur (2006):
the torus could be made of individual clouds, with the "X-ray
torus" being different from the "dusty torus", instead of a
continuous absorbing medium with sharp edges.

As shown at the beginning of the paper, we are biased against
faint optical counterparts ($R\geq 22$). It would thus be very
interesting to extend our work to significantly fainter optical
counterparts in order to check whether the fraction of X-ray
sources for which the X-ray and the optical classifications do not
match is still that large.

\section{ACKNOWLEDGMENTS}
We thank the referee for useful suggestions. We are indebted to D.
Hutsem\'{e}kers for some fruitful discussions. This work is based
on observations obtained with XMM-Newton, an ESA science mission
with instruments and contributions directly funded by ESA Member
States and NASA. OG, EG, PGS, VB, and JS acknowledge support from
the ESA PRODEX Programme "XMM-LSS", and from the Belgian Federal
Science Policy Office for their support. LC, DM, MT acknowledge
financial contribution from contract ASI-INAF I/023/05/0. PG is a
Fellow of the Japan Society for the Promotion of Science. MK
acknowledges the German DLR under grant 50OR0404. ACR, JEC and FJC
acknowledge financial support from the Spanish Ministry for
Education and Science, under projects ESP2003-00812 and
ESP2006-13608-C02-01. JAT, SM, SR and MW acknowledge support from
the UK PPARC research council.

\Online

\setcounter{table}{3}

\begin{sidewaystable*}[ht!]
\caption[]{Optical properties and best-fit parameters from the
X-ray spectral analysis.}
\begin{tabular}{ccccccccccccccc}\hline \hline
\# & Name & RA & DECL & $HR$  & X-type & OPT-type & $R$ &
$F$$_{2-10}$ & Model & $z_{\mathrm{spec}}$ & log
$L_{2-10}^{\mathrm{corr}}$ & $N_{\mathrm{H}}$
& $\Gamma$ & $\chi^{2}_{\nu}$ \\
& & (deg) & (deg) & & & & & (10$^{-14}$ cgs) & & & & (10$^{22}$ cm$^{-2}$)& & \\
(1) & (2) & (3) & (4) & (5) & (6) & (7) & (8) & (9) & (10) & (11)
& (12) & (13) & (14) & (15)\\ \hline
1 & XLSS J022100.2$-$042855.2 & 02:21:00.11  &  $-$4:28:54.93  &  $-$0.66  &  I & 2 & 18.4 & 1.43  &  PL  &  0.200  &  42.27  &  $0.026^{+0.025}_{-0.000}$ & $2.57_{-0.21}^{+0.24}$ & 1.24 \\
2 & XLSS J022202.7$-$050944.3 & 02:22:02.68  &  $-$5:09:44.42  &  $-$0.39  &  I & 1 & 18.4 & 24.60  &  PL  &  0.275  &  43.73  &  $0.026^{+0.052}_{-0.000}$ & $1.66_{-0.14}^{+0.08}$ & 0.61 \\
3 & XLSS J022208.2$-$042732.2 & 02:22:08.22  &  $-$4:27:33.22  &  $-$0.59  &  I & 1 & 18.4 & 2.45  &  PL  &  1.678  &  44.64  &  $0.026^{+0.214}_{-0.000}$ & 1.9 & - \\
4 & XLSS J022234.4$-$043707.6 & 02:22:34.43  &  $-$4:37:08.52  &  $-$0.42  &  I & 2 & 19.6 & 1.28  &  PL  &  0.167  &  42.00  &  $0.085^{+0.335}_{-0.059}$ & 1.9 & - \\
5 & XLSS J022244.4$-$043346.8 & 02:22:44.40  &  $-$4:33:46.35  &  $-$0.51  &  I & 1 & 17.9 & 21.76  &  PL  &  0.760  &  44.77  &  0.026 & $1.99_{-0.04}^{+0.04}$ & 0.96\\
6 & XLSS J022247.8$-$043329.1 & 02:22:47.78  &  $-$4:33:30.26  &  $-$0.58  &  I & 1 & 19.7 & 1.57  &  PL  &  1.629  &  44.40  &  $0.026^{+0.289}_{-0.000}$ & $1.85_{-0.25}^{+0.26}$ & 1.76 \\
7 & XLSS J022249.3$-$051452.1 & 02:22:49.51  &  $-$5:14:52.64  &  $-$0.64  &  I & 1 & 17.9 & 6.02  &  PL  &  0.313  &  43.34  &  $0.026^{+0.011}_{-0.000}$ & $2.42_{-0.09}^{+0.09}$ & 1.02 \\
8 & XLSS J022249.6$-$041350.0 & 02:22:49.58  &  $-$4:13:52.22  &  $-$0.55  &  I & 1 & 20.1 & 1.43  &  PL  &  1.566  &  44.46  &  $0.026^{+0.147}_{-0.000}$ & $2.15_{-0.21}^{+0.24}$ & 1.16 \\
9 & XLSS J022253.6$-$042931.3 & 02:22:53.58  &  $-$4:29:27.93  &  $-$0.40  &  I & 2 & 18.1 & 3.58  &  PL  &  0.215  &  42.68  &  $0.026^{+0.053}_{-0.000}$ & 1.9 & 1.01 \\
10 & XLSS J022258.0$-$041840.3 & 02:22:57.97  &  $-$4:18:40.40  &  0.57  &  II & 2 & 17.5 & 12.23  &  PL+GA$^a$  &  0.237  &  43.37  &  $3.43_{-1.30}^{+5.30}$ & 1.9 & 0.72 \\
11 & XLSS J022301.9$-$043205.2 & 02:23:01.93  &  $-$4:32:04.65  &  $-$0.29  &  II & 2 & 19.7 & 0.97  &  PL  &  0.616  &  43.20  &  $1.390_{-0.808}^{+1.202}$ & 1.9 & - \\
12 & XLSS J022312.1$-$050624.9 & 02:23:12.26  &  $-$5:06:25.45  &  $-$0.41  &  I & 1 & 19.4 & 1.30  &  PL  &  2.206  &  44.65  & $0.217_{-0.191}^{+1.361}$  & 1.9 & - \\
13 & XLSS J022313.3$-$043102.6 & 02:23:13.20  &  $-$4:31:00.93  &  $-$0.35  &  I & 2 & 15.6 & 1.23  &  PL  &  0.017  &  39.90  &  $0.026_{-0.000}^{+0.090}$ & 1.9 & - \\
14 & XLSS J022317.2$-$044032.4 & 02:23:17.34  &  $-$4:40:33.46  &  $-$0.44  &  I & 1 & 18.0 & 3.62  &  PL  &  0.842  &  43.96  &  $0.026_{-0.000}^{+0.151}$ & $1.34_{-0.22}^{+0.27}$ & 1.20 \\
15 & XLSS J022319.5$-$044732.1 & 02:23:19.54  &  $-$4:47:31.50  &  $-$0.40  &  I & 2 & 18.6 & 2.72  &  PL  &  0.293  &  42.87  &  $0.203_{-0.126}^{+0.116}$ & $1.87_{-0.22}^{+0.28}$ & 1.00 \\
16 & XLSS J022320.9$-$050629.0 & 02:23:21.05  &  $-$5:06:27.20  &  $-$0.42  &  I & 1 & 19.7 & 2.03  &  PL  &  2.415  &  44.94  &  $0.200_{-0.174}^{+0.580}$ & 1.9 & - \\
17 & XLSS J022326.5$-$045706.9 & 02:23:26.45  &  $-$4:57:04.58  &  $-$0.32  &  I & 1 & 20.5 & 1.52  &  PL  &  0.826  &  43.68  &  0.026 & $1.89_{-0.46}^{+0.51}$ & - \\
18 & XLSS J022327.8$-$040119.0 & 02:23:27.85  &  $-$4:01:19.71  &  $-$0.68  &  I & 1 & 19.7 & 1.80  &  PL  &  1.922  &  44.73  &  0.026 & $2.08_{-0.20}^{+0.22}$ & 1.48 \\
19 & XLSS J022329.3$-$045453.0 & 02:23:29.18  &  $-$4:54:53.12  &  $-$0.66  &  I & 1 & 19.6 & 1.58  &  PL  &  0.604  &  43.45  &  $0.026^{+0.041}_{-0.000}$ & $2.26_{-0.19}^{+0.25}$ & 0.73 \\
20 & XLSS J022330.0$-$043002.4 & 02:23:30.08  &  $-$4:30:01.38  &  $-$0.39  &  II & 1 & 20.0 & 2.35  &  PL  &  2.666  &  45.11  &  $4.002_{-2.008}^{+2.901}$ & 1.9 & - \\
21 & XLSS J022333.2$-$044925.1 & 02:23:33.14  &  $-$4:49:24.79  &  $-$0.54  &  I & 1 & 20.6 & 1.15  &  PL  &  2.302  &  44.65  &  $0.173^{+0.819}_{-0.147}$ & 1.9 & - \\
22 & XLSS J022337.6$-$044006.3 & 02:23:37.54  &  $-$4:40:06.31  &  $-$0.63  &  I & 1 & 19.6 & 0.53  &  PL  &  1.743  &  44.23  &  0.026 & $2.38_{-0.18}^{+0.39}$ & - \\
23 & XLSS J022347.1$-$050551.6 & 02:23:47.22  &  $-$5:05:51.99  &  $-$0.51  &  I & 1 & 19.4 & 1.35  &  PL  &  2.329  &  44.71  &  0.026 & $1.88_{-0.26}^{+0.21}$ & 0.86 \\
24 & XLSS J022351.2$-$042053.8 & 02:23:51.18  &  $-$4:20:54.33  &  0.14  &  II & 2 & 19.2 & 2.94  &  PL  &  0.181  &  42.49  &  $2.382_{-1.424}^{+3.114}$ & 1.9 & 0.82 \\
\hline

\end{tabular}

Column 1 : internal sequence number;

Column 2 : source name, using the convention defined in Pierre et
al. (2007);

Column 3 : right ascension of the X-ray source coming from the
band-merged catalog from the [0.5-2] and [2-10] keV bands, which
has been astrometrically corrected (Pierre et al. 2007);

Column 4 : declination of the X-ray source coming from the
band-merged catalog from the [0.5-2] and [2-10] keV bands, which
has been astrometrically corrected (Pierre et al. 2007);

Column 5 : hardness ratio between the hard (H) [2-10] keV band and
the soft (S) [0.5-2] keV band computed using the formula
$\frac{H-S}{H+S}$;

Column 6 : X-ray classification based on the X-ray spectral
analysis : II (N$_{H}$$>$ 10$^{22}$ cm$^{-2}$) and I (N$_{H}$$<$
10$^{22}$ cm$^{-2}$);

Column 7 : optical classification based on the optical spectra: 1
(FWHM$>$1500 km s$^{-1}$) and 2 (FWHM$<$1500 km s$^{-1}$);

Column 8 : $R$ band magnitude from the SuperCosmos Sky Survey or
from the VVDS;

Column 9 : [2-10] keV band flux of the X-ray source derived from
the fitted X-ray model;

Column 10 : best-fit model (PL= absorbed power law; PL+BB=
absorbed power law plus black body; PL+POW= absorbed power law
plus unabsorbed power law; PL+GA= absorbed power law plus an
additional emission line;

Column 11 : spectroscopic redshift derived from the optical
spectra;

Column 12 : logarithm of the intrinsic (corrected for absorption,
for both the galactic and the intrinsic column density) rest-frame
luminosity in the [2-10] keV band [erg s$^{-1}$];

Column 13 : best-fit hydrogen column density [10$^{22}$ cm$^{-2}$]
along with its error bar which represents a 90\% confidence
interval. The value 0.026 corresponds to the galactic value and
has been fixed for the values without error bars;

Column 14 : best-fit value for the photon index $\Gamma$ along
with its error bars which are for a 90\% confidence interval. For
some objects we fixed $\Gamma$ to a value of 1.9 which is
representative of a type 1 AGN.

Column 15 : statistic which has been used. We present the value
for the reduced chi square $\chi^{2}_{\nu}$. Otherwise, Cash
statistic has been used.

$^a$ The presence of an emission line around 4.0 keV (observed
frame) is required.

$^b$ The spectroscopic redshift is tentative.

$^c$ The presence of an emission line around 0.9 keV (observed
frame) is required, but see text.

$^d$ The presence of an emission line at 6.4 keV (source rest
frame, consistent with Fe K) is required.

$^e$ The presence of an emission line around 0.7 keV (observed
frame) is required, but see text.

$^f$ The presence of an emission line at 6.4 keV (source rest
frame, consistent with Fe K) is required.

 \label{sample.tbl}
\end{sidewaystable*}

\setcounter{table}{3}

\begin{sidewaystable*}[ht!]
\caption[]{Continued.}

\footnotesize
\begin{tabular}{ccccccccccccccc}\hline \hline
\# & Name &RA & DECL & $HR$  & X-type & OPT-type & $R$ &
$F_{2-10}$ & Model & $z_{\mathrm{spec}}$ & log
$L_{2-10}^{\mathrm{corr}}$ & $N_{\mathrm{H}}$
& $\Gamma$ & $\chi^{2}_{\nu}$ \\
& & (deg) & (deg) & & & & & (10$^{-14}$ cgs) & & & & (10$^{22}$ cm$^{-2}$)& & \\
(1) & (2) & (3) & (4) & (5) & (6) & (7) & (8) & (9) & (10) & (11)
& (12) & (13)& (14) & (15)\\ \hline
25 & XLSS J022354.7$-$044818.1 & 02:23:54.67  &  $-$4:48:14.32  &  $-$0.67  &  I & 1 & 17.6 & 0.77  &  PL  &  2.458  &  44.85  &  0.026 & $2.48_{-0.24}^{+0.26}$ & 0.86 \\
26 & XLSS J022402.0$-$025256.1 & 02:24:02.02  &  $-$2:52:54.03  &  $-$0.44  &  I & 2 & 17.7 & 3.96  &  PL  &  0.102  &  42.00  &  0.026 & $1.51_{-0.32}^{+0.33}$ & - \\
27 & XLSS J022402.5$-$051343.3 & 02:24:02.59  &  $-$5:13:43.31  &  0.23  &  II & 2 & 17.7 & 2.65  &  PL+PL  &  0.084  &  42.32  &  $51.245_{-26.168}^{+57.417}$ & 1.9 & - \\
28 & XLSS J022410.9$-$050656.7 & 02:24:10.87  &  $-$5:06:54.01  &  $-$0.51  &  I & 1 & 19.9 & 1.69  &  PL  &  2.314  &  44.86  &  0.026 & $1.99_{-0.26}^{+0.29}$ & 1.62 \\
29 & XLSS J022413.7$-$032918.7 & 02:24:13.72  &  $-$3:29:18.73  &  0.05  &  II & 1 & 17.6 & 8.61  &  PL+BB  &  0.429  &  43.79  &  $2.015_{-1.02755}^{+1.18785}$ & 1.9 & 0.61 \\
30 & XLSS J022435.8$-$030849.3 & 02:24:35.86  &  $-$3:08:52.05  &  $-$0.46  &  I & 1 & 20.4 & 2.94  &  PL  &  1.946  &  44.87  &  0.026 & $1.90_{-0.26}^{+0.28}$ & - \\
31$^b$ & XLSS J022438.5$-$051150.4 & 02:24:38.59  &  $-$5:11:48.07  &  $-$0.53  &  II & 1 & -1 & 1.38  &  PL  &  2.283  &  44.71  & $1.360_{-0.913}^{+1.091}$  & 1.9 & - \\
32 & XLSS J022438.9$-$033305.6 & 02:24:38.92  &  $-$3:33:05.63  &  $-$0.74  &  I & 1 & 18.1 & 2.76  &  PL  &  1.676  &  44.81  &  0.026 & $2.18_{-0.19}^{+0.20}$ & 1.18 \\
33 & XLSS J022439.7$-$042402.2 & 02:24:39.71  &  $-$4:23:59.07  &  $-$0.47  &  I & 2 & 19.8 & 3.11  &  PL  &  0.478  &  43.40  &  $0.026^{+0.025}_{-0.000}$ & $1.78_{-0.23}^{+0.24}$ & 1.05 \\
34 & XLSS J022440.7$-$043656.8 & 02:24:40.72  &  $-$4:36:57.40  &  $-$0.59  &  I & 1 & 19.7 & 0.86  &  PL  &  0.904  &  43.68  &  0.026 & $2.41_{-0.27}^{+0.29}$ & - \\
35 & XLSS J022446.9$-$030344.1 & 02:24:46.84  &  $-$3:03:46.93  &  $-$0.51  &  I & 2 & 18.5 & 4.73  &  PL  &  0.301  &  43.13  &  $0.026^{+0.067}_{-0.000}$ & 1.9 & - \\
36 & XLSS J022507.4$-$043406.3 & 02:25:07.32  &  $-$4:34:04.00  &  $-$0.46  &  I & 2 & 21.6 & 1.11  &  PL  &  0.617  &  43.21  &  0.026 & $1.74_{-0.45}^{+0.50}$ & - \\
37 & XLSS J022514.4$-$044658.8 & 02:25:14.30  &  $-$4:46:58.52  &  $-$0.61  &  I & 1 & 18.0 & 1.81  &  PL  &  1.924  &  44.84  &  0.026 & $2.29_{-0.21}^{+0.23}$ & 1.08 \\
38 & XLSS J022519.2$-$044714.3 & 02:25:19.34  &  $-$4:47:12.91  &  $-$0.47  &  I & 1 & 20.9 & 1.06  &  PL  &  1.929  &  44.49  &  0.026 & $2.04_{-0.42}^{+0.47}$ & - \\
39 & XLSS J022521.1$-$043947.9 & 02:25:21.11  &  $-$4:39:48.98  &  $-$0.12  &  I & 2 & 18.8 & 4.55  &  PL  &  0.265  &  43.00  &  $0.524_{-0.226}^{+0.311}$ & 1.9 & 0.93 \\
40 & XLSS J022521.2$-$032627.1 & 02:25:21.29  &  $-$3:26:28.73  &  $-$0.50  &  I & 1 & 20.4 & 2.05  &  PL  &  1.067  &  44.17  &  0.026 & $2.16_{-0.21}^{+0.22}$ & - \\
41 & XLSS J022521.4$-$035210.0 & 02:25:21.32  &  $-$3:52:09.25  &  $-$0.42  &  I & 2 & 17.4 & 4.85  &  PL  &  0.098  &  42.07 &  $0.026^{+0.01}_{-0.000}$ & $2.01_{-0.15}^{+0.15}$ & - \\
42 & XLSS J022522.9$-$042649.5 & 02:25:22.91  &  $-$4:26:49.05  &  0.22  &  II & 2 & 21.8 & 2.12  &  PL  &  1.029  &  44.10  &  $5.349_{-1.990}^{+2.853}$ & 1.9 & - \\
43 & XLSS J022524.5$-$025829.2 & 02:25:24.53  &  $-$2:58:29.63  &  $-$0.64  &  I & 1 & 18.9 & 5.29  &  PL  &  1.229  &  44.65  &  $0.026^{+0.069}_{-0.000}$ & 1.9 & - \\
44 & XLSS J022524.5$-$044043.1 & 02:25:24.82  &  $-$4:40:43.61  &  $-$0.80  &  I & glx & 17.3 & 3.96  &  PL+GA$^c$  &  0.263  &  42.92  &  $0.026^{+0.029}_{-0.000}$ & 1.9 & 1.34\\
45 & XLSS J022525.3$-$034225.4 & 02:25:25.28  &  $-$3:42:24.26  &  $-$0.66  &  I & 1 & 17.5 & 5.97  &  PL  &  1.326  &  44.85  &  0.026 & $2.09_{-0.17}^{+0.17}$ & 0.87 \\
46 & XLSS J022527.6$-$035954.9 & 02:25:27.52  &  $-$3:59:53.89  &  $-$0.50  &  I & 1 & 20.6 & 2.69  &  PL  &  0.494  &  43.36  &  0.026 & $1.69_{-0.34}^{+0.37}$ & - \\
47 & XLSS J022527.6$-$050639.2 & 02:25:27.62  &  $-$5:06:39.19  &  $-$0.48  &  I & 1 & 20.1 & 2.43  &  PL  &  1.114  &  44.26  &  $0.227_{-0.227}^{+0.402}$ & $2.08_{-0.36}^{+0.29}$ & 0.63 \\
48 & XLSS J022529.4$-$050947.8 & 02:25:29.32  &  $-$5:09:46.95  &  $-$0.03  &  II & 1 & 19.3 & 2.35  &  PL  &  1.288  &  44.37  &  $5.226_{-2.229}^{+2.927}$ & 1.9 & - \\
\hline

\end{tabular}
\end{sidewaystable*}

 \setcounter{table}{3}

\begin{sidewaystable*}[ht!]
\caption[]{Continued.}

\footnotesize
\begin{tabular}{ccccccccccccccc}\hline \hline
\#& Name &RA & DECL & $HR$  & X-type & OPT-type & $R$ & $F_{2-10}$
& Model & $z_{\mathrm{spec}}$ & log $L_{2-10}^{\mathrm{corr}}$ &
$N_{\mathrm{H}}$
& $\Gamma$ & $\chi^{2}_{\nu}$ \\
& & (deg) & (deg) & & & & & (10$^{-14}$ cgs) & & & & (10$^{22}$ cm$^{-2}$)& & \\
(1) & (2) & (3) & (4) & (5) & (6) & (7) & (8) & (9) & (10) & (11)
& (12) & (13)& (14) & (15) \\ \hline
49 & XLSS J022533.6$-$050803.1 & 02:25:33.67  &  $-$5:08:02.44  &  $-$0.67  &  I & 1 & 17.5 & 2.40  &  PL  &  1.179  &  44.26  &  $0.026^{+0.032}_{-0.000}$ & 1.9 & - \\
50 & XLSS J022537.0$-$050110.7 & 02:25:37.06  &  $-$5:01:09.38  &  $-$0.47  &  I & 1 & 19.8 & 1.71  &  PL  &  1.937  &  44.72  &  0.026 & $2.08_{-0.23}^{+0.25}$ & 0.60 \\
51 & XLSS J022538.3$-$042109.8 & 02:25:38.32  &  $-$4:21:09.32  &  $-$0.48  &  I & 1 & 20.9 & 1.28  &  PL  &  0.742  &  43.50  &  0.026 & $1.90_{-0.25}^{+0.56}$ & - \\
52 & XLSS J022541.0$-$033521.9 & 02:25:40.84  &  $-$3:35:21.41  &  $-$0.59  &  I & 1 & 18.3 & 2.27  &  PL  &  0.861  &  43.90  &  $0.026^{+0.025}_{-0.000}$ & 1.9 & - \\
53 & XLSS J022541.4$-$032624.2 & 02:25:41.30  &  $-$3:26:24.12  &  $-$0.31  &  I & 2 & 18.5 & 12.93  &  PL  &  0.207  &  43.21  &  $0.253_{-0.068}^{+0.079}$ & 1.9 & 0.99 \\
54 & XLSS J022544.8$-$043736.7 & 02:25:44.98  &  $-$4:37:36.36  &  $-$0.35  &  II & 1 & 22.6 & 1.01  &  PL  &  3.589  &  45.04  & $3.501_{-2.538}^{+2.445}$  & 1.9 & - \\
55 & XLSS J022546.3$-$051159.1 & 02:25:46.38  &  $-$5:12:02.52  &  $-$0.42  &  I & glx & 17.0 & 1.55  &  PL  &  0.148  &  41.96  &  $0.026^{+0.060}_{-0.000}$ & $1.83_{-0.24}^{+0.36}$ & 0.93 \\
56 & XLSS J022548.7$-$025820.2 & 02:25:48.72  &  $-$2:58:20.02  &  $-$0.53  &  I & 2 & 17.7 & 10.27  &  PL  &  0.286  &  43.44  &  $0.026^{+0.019}_{-0.000}$ & $1.93_{-0.09}^{+0.10}$ & 1.76\\
57 & XLSS J022554.8$-$051355.5 & 02:25:54.84  &  $-$5:13:55.51  &  $-$0.38  &  II & 1 & 20.0 & 1.00  &  PL  &  1.247  &  43.95  &  $1.566_{-1.200}^{+2.202}$ & 1.9 & - \\
58 & XLSS J022556.1$-$044725.2 & 02:25:56.14  &  $-$4:47:24.13  &  $-$0.50  &  I & 1 & 20.0 & 3.51  &  PL  &  1.010  &  44.21  &  $0.026$ & $1.73_{-0.27}^{+0.20}$ & 0.88 \\
59 & XLSS J022556.3$-$042540.9 & 02:25:56.28  &  $-$4:25:40.47  &  $-$0.69  &  I & 2 & 21.5 & 1.13  &  PL  &  0.863  &  43.62  &  0.026 & $1.97_{-0.34}^{+0.37}$ & - \\
60 & XLSS J022557.6$-$045005.9 & 02:25:57.65  &  $-$4:50:04.42  &  $-$0.53  &  I & 1 & 19.3 & 0.47  &  PL  &  2.263  &  44.33  &  0.026 & $2.08_{-0.41}^{+0.45}$ & - \\
61 & XLSS J022558.7$-$050055.8 & 02:25:58.73  &  $-$5:00:55.24  &  0.62  &  II & 2 & 17.2 & 5.91  &  PL  &  0.148  &  42.80  &  $12.985_{-5.945}^{+11.938}$ & 1.9 & - \\
62 & XLSS J022601.5$-$033041.0 & 02:26:01.57  &  $-$3:30:39.66  &  $-$0.49  &  I & 1 & 19.8 & 1.21  &  PL  &  1.455  &  44.19  &  $0.026^{+0.167}_{-0.000}$ & 1.9 & - \\
63 & XLSS J022604.5$-$045934.2 & 02:26:04.31  &  $-$4:59:29.48  &  $-$0.23  &  I,Fe & 2 & 15.4 & 2.56  &  PL+GA$^d$  &  0.054  &  41.25  &  $0.026^{+0.024}_{-0.000}$ & 1.9 & - \\
64 & XLSS J022607.1$-$032015.0 & 02:26:07.15  &  $-$3:20:15.02  &  $-$0.45  &  I & 1 & 19.1 & 1.88  &  PL  &  2.321  &  44.86  &  $0.026_{-0.000}^{+1.275}$ & 1.9 & - \\
65 & XLSS J022607.8$-$041842.7 & 02:26:07.76  &  $-$4:18:42.07  &  $-$0.61  &  I & 1 & 18.9 & 4.41  &  PL  &  0.495  &  43.72  &  $0.026^{+0.023}_{-0.000}$ & $2.48_{-0.18}^{+0.23}$ & 1.18 \\
66 & XLSS J022617.2$-$044725.3 & 02:26:17.23  &  $-$4:47:24.83  &  0.33  &  II & 2 & 17.4 & 2.71  &  PL  &  0.140  &  42.25  &  $3.717_{-1.851}^{+2.572}$ & 1.9 & - \\
67 & XLSS J022617.3$-$050443.8 & 02:26:17.63  &  $-$5:04:43.83  &  $-$0.51  &  II & 2 & 14.8 & 4.27  &  PL+GA$^e$  &  0.054  &  41.56  &  $2.680_{-1.952}^{+116.01}$ & 1.9 & - \\
68 & XLSS J022617.7$-$043107.6 & 02:26:17.74  &  $-$4:31:07.09  &  $-$0.78  &  I & 1 & 18.7 & 1.07  &  PL  &  0.705  &  43.53  &  0.026 & $2.63_{-0.27}^{+0.32}$ & 1.23 \\
69 & XLSS J022620.5$-$031723.9 & 02:26:20.72  &  $-$3:17:26.93  &  $-$0.19  &  I,Fe & 1 & 19.2 & 7.61  &  PL+GA$^f$  &  0.691  &  44.21  &  $0.677_{-0.504}^{+0.595}$ & 1.9 & - \\
70 & XLSS J022622.3$-$042219.9 & 02:26:22.09  &  $-$4:22:20.30  &  $-$0.52  &  II & 1 & 18.5 & 5.04  &  PL  &  2.011  &  45.15  &  $1.242_{-0.917}^{+1.071}$ & 1.9 & - \\
71 & XLSS J022629.2$-$043056.4 & 02:26:29.18  &  $-$4:30:56.24  &  $-$0.43  &  I & 1 & 19.8 & 3.96  &  PL  &  2.031  &  45.05  & $0.026_{-0.000}^{+1.426}$  & 1.9 & - \\
72 & XLSS J022643.6$-$043317.1 & 02:26:43.62  &  $-$4:33:16.57  &  0.58  &  II & glx & 17.8 & 2.52  &  PL  &  0.308  &  42.96  &  $3.861_{-2.211}^{+3.607}$ & 1.9 & - \\
73 & XLSS J022646.2$-$045156.4 & 02:26:46.25  &  $-$4:51:55.91  &  $-$0.61  &  I & 1 & 20.9 & 0.63  &  PL  &  1.089  &  43.73  &  0.026 & $2.32_{-0.38}^{+0.44}$ & - \\
\hline

\end{tabular}
\end{sidewaystable*}

\setcounter{table}{3}

\begin{sidewaystable*}[ht!]
\caption[]{Continued.}
\footnotesize
\begin{tabular}{ccccccccccccccc}\hline \hline
\#& Name &RA & DECL & $HR$  & X-type & OPT-type & $R$ & $F_{2-10}$
& Model & $z_{\mathrm{spec}}$ & log $L_{2-10}^{\mathrm{corr}}$ &
$N_{\mathrm{H}}$
& $\Gamma$ & $\chi^{2}_{\nu}$ \\
& &(deg) & (deg) & & & & & (10$^{-14}$ cgs) & & & &(10$^{22}$ cm$^{-2}$) & & \\
(1) & (2) & (3) & (4) & (5) & (6) & (7) & (8) & (9) & (10) & (11)
& (12) & (13)& (14)& (15) \\ \hline
74 & XLSS J022646.9$-$041839.6 & 02:26:46.93  &  $-$4:18:39.15  &  $-$0.52  &  I & 1 & 21.0 & 1.20  &  PL  &  1.581  &  44.27  &  $0.026^{+0.682}_{-0.000}$ & 1.9 & - \\
75 & XLSS J022649.0$-$042745.7 & 02:26:48.95  &  $-$4:27:45.74  &  $-$0.56  &  I & 2 & 19.0 & 2.52  &  PL  &  0.327  &  42.94  &  $0.026^{+0.080}_{-0.000}$ & 1.9 & - \\
76 & XLSS J022649.4$-$041155.4 & 02:26:49.38  &  $-$4:11:53.79  &  $-$0.55  &  I & 1 & 22.2 & 2.19  &  PL  &  1.157  &  44.15  &  0.026 & $1.75_{-0.19}^{+0.20}$ & 0.95 \\
77 & XLSS J022651.7$-$045715.3 & 02:26:51.65  &  $-$4:57:13.34  &  $-$0.49  &  I & 2 & 20.4 & 3.85  &  PL  &  0.331  &  43.12  &  $0.026$ & $1.72_{-0.21}^{+0.23}$ & 0.65 \\
78 & XLSS J022652.6$-$050619.6 & 02:26:52.44  &  $-$5:06:20.18  &  $-$0.49  &  I & 1 & 18.7 & 3.82  &  PL  &  1.618  &  44.76  &  0.026 & $1.83_{-0.26}^{+0.27}$ & - \\
79 & XLSS J022659.8$-$044430.4 & 02:26:59.78  &  $-$4:44:29.86  &  $-$0.54  &  I & 1 & 20.8 & 0.45  &  PL  &  1.612  &  43.97  &  0.026 & $2.15_{-0.44}^{+0.54}$ & - \\
80 & XLSS J022701.3$-$040751.2 & 02:27:01.40  &  $-$4:07:49.95  &  $-$0.35  &  I & 2 & 18.6 & 4.24  &  PL  &  0.220  &  42.75  &  $0.026^{+0.038}_{-0.000}$ & $1.56_{-0.14}^{+0.16}$ & 1.14 \\
81 & XLSS J022707.8$-$050816.5 & 02:27:07.78  &  $-$5:08:15.97  &  0.32  &  II & 2 & 18.9 & 16.43  &  PL  &  0.356  &  43.90  &  $2.906_{-0.682}^{+0.779}$ & 1.9 & - \\
82 & XLSS J022708.5$-$050429.1 & 02:27:08.60  &  $-$5:04:23.92  &  $-$0.48  &  I & 2 & 17.6 & 4.69  &  PL  &  0.148  &  42.30  &  0.026 & $1.92_{-0.25}^{+0.27}$ & - \\
83 & XLSS J022712.0$-$033353.1 & 02:27:11.84  &  $-$3:33:51.23  &  $-$0.40  &  I & 2 & 18.5 & 2.70  &  PL  &  0.317  &  42.95  &  $0.026^{+0.075}_{-0.000}$ & $1.99_{-0.26}^{+0.34}$ & 0.78 \\
84 & XLSS J022712.7$-$044220.7 & 02:27:12.67  &  $-$4:42:20.25  &  0.14  &  II & 2 & 18.4 & 4.69  &  PL  &  0.205  &  42.79  &  $1.461_{-0.582}^{+0.777}$ & 1.9 & 1.01 \\
85 & XLSS J022729.3$-$043226.9 & 02:27:29.48  &  $-$4:32:24.14  &  $-$0.51  &  I & 1 & 19.6 & 1.52  &  PL  &  2.290  &  44.76  &  0.026 & $1.89_{-0.23}^{+0.24}$ & - \\
86 & XLSS J022731.9$-$032121.9 & 02:27:31.90  &  $-$3:21:22.13  &  $-$0.62  &  I & 1 & -1 & 0.53  &  PL  &  0.771  &  43.32  &  0.026 & $2.58_{-0.67}^{+0.89}$ & - \\
87 & XLSS J022732.2$-$032736.0 & 02:27:32.26  &  $-$3:27:36.29  &  $-$0.38  &  I & 1 & 18.0 & 7.72  &  PL  &  1.782  &  45.11  &  $0.026_{-0.000}^{+0.578}$ & $1.69_{-0.18}^{+0.27}$ & 0.92 \\
88 & XLSS J022735.7$-$042023.2 & 02:27:35.75  &  $-$4:20:22.69  &  $-$0.55  &  I & 2 & -1 & 1.21  &  PL  &  0.886  &  43.66  &  $0.150_{-0.124}^{+0.193}$ & 1.9 & - \\
89 & XLSS J022739.7$-$050043.4 & 02:27:39.85  &  $-$5:00:43.16  &  $-$0.61  &  I & 1 & 18.8 & 7.75  &  PL  &  0.447  &  43.80  &  $0.026_{-0.000}^{+0.011}$ & $2.18_{-0.09}^{+0.13}$ & 1.03 \\
90 & XLSS J022740.4$-$041855.5 & 02:27:40.68  &  $-$4:18:57.91  &  $-$0.17  &  I & 1 & 22.4 & 0.71  &  PL  &  0.727  &  43.25  &  0.026 & $2.016_{-0.71}^{+0.94}$ & - \\
91 & XLSS J022748.4$-$041900.8 & 02:27:48.53  &  $-$4:19:00.82  &  $-$0.32  &  I & 2 & 21.4 & 1.15  &  PL  &  0.689  &  43.32  &  0.026 & $1.67_{-0.31}^{+0.34}$  & - \\
92 & XLSS J022751.1$-$050101.2 & 02:27:51.34  &  $-$5:01:01.73  &  $-$0.47  &  I & 1 & -1 & 3.36  &  PL  &  2.174  &  45.01  &  0.026 & $1.83_{-0.26}^{+0.21}$ & 0.47 \\
93 & XLSS J022754.5$-$050700.1 & 02:27:54.43  &  $-$5:06:58.27  &  $-$0.44  &  I & 2 & 19.5 & 1.46  &  PL  &  0.436  &  43.00  &  $0.262^{+0.204}_{-0.152}$ & 1.9 & - \\
94 & XLSS J022756.8$-$050734.8 & 02:27:56.92  &  $-$5:07:34.97  &  $-$0.13  &  II & 2 & 19.2 & 1.82  &  PL  &  0.492  &  43.23  &  $1.085_{-0.536}^{+0.669}$ & 1.9 & - \\
95 & XLSS J022809.1$-$041233.5 & 02:28:09.01  &  $-$4:12:31.12  &  $-$0.49  &  I & 1 & 19.2 & 25.55  &  PL  &  0.878  &  44.91  &  0.026 & $1.67_{-0.18}^{+0.19}$ & 1.86 \\
96 & XLSS J022810.6$-$050744.5 & 02:28:10.81  &  $-$5:07:43.84  &  $-$0.61  &  I & 1 & -1 & 0.84  &  PL  &  0.733  &  43.30  &  $0.026^{+0.026}_{-0.000}$ & 1.9 & - \\
97 & XLSS J022821.4$-$050954.8 & 02:28:21.54  &  $-$5:09:55.88  &  $-$0.63  &  I & 1 & 20.7 & 1.01  &  PL  &  1.812  &  44.43  &  0.026 & $2.11_{-0.29}^{+0.31}$ & - \\
98 & XLSS J022843.3$-$051011.5 & 02:28:43.36  &  $-$5:10:11.61  &  $-$0.02  &  II & 2 & 18.5 & 2.91  &  PL  &  0.271  &  42.64  &  $1.652_{-0.791}^{+1.370}$ & 1.9 & - \\
99 & XLSS J022851.3$-$051223.0 & 02:28:51.42  &  $-$5:12:23.07  &  $-$0.72  &  I & 1 & 17.1 & 37.17  &  PL  &  0.315  &  44.10  &  $0.026^{+0.008}_{-0.000}$ & $2.11_{-0.05}^{+0.05}$ & 1.12 \\
\hline

\end{tabular}
\end{sidewaystable*}

\clearpage

\appendix

\section{Illustrating figures about selection effects in our sample}

\begin{figure*}
\centering
  \includegraphics[angle=0,width=18cm]{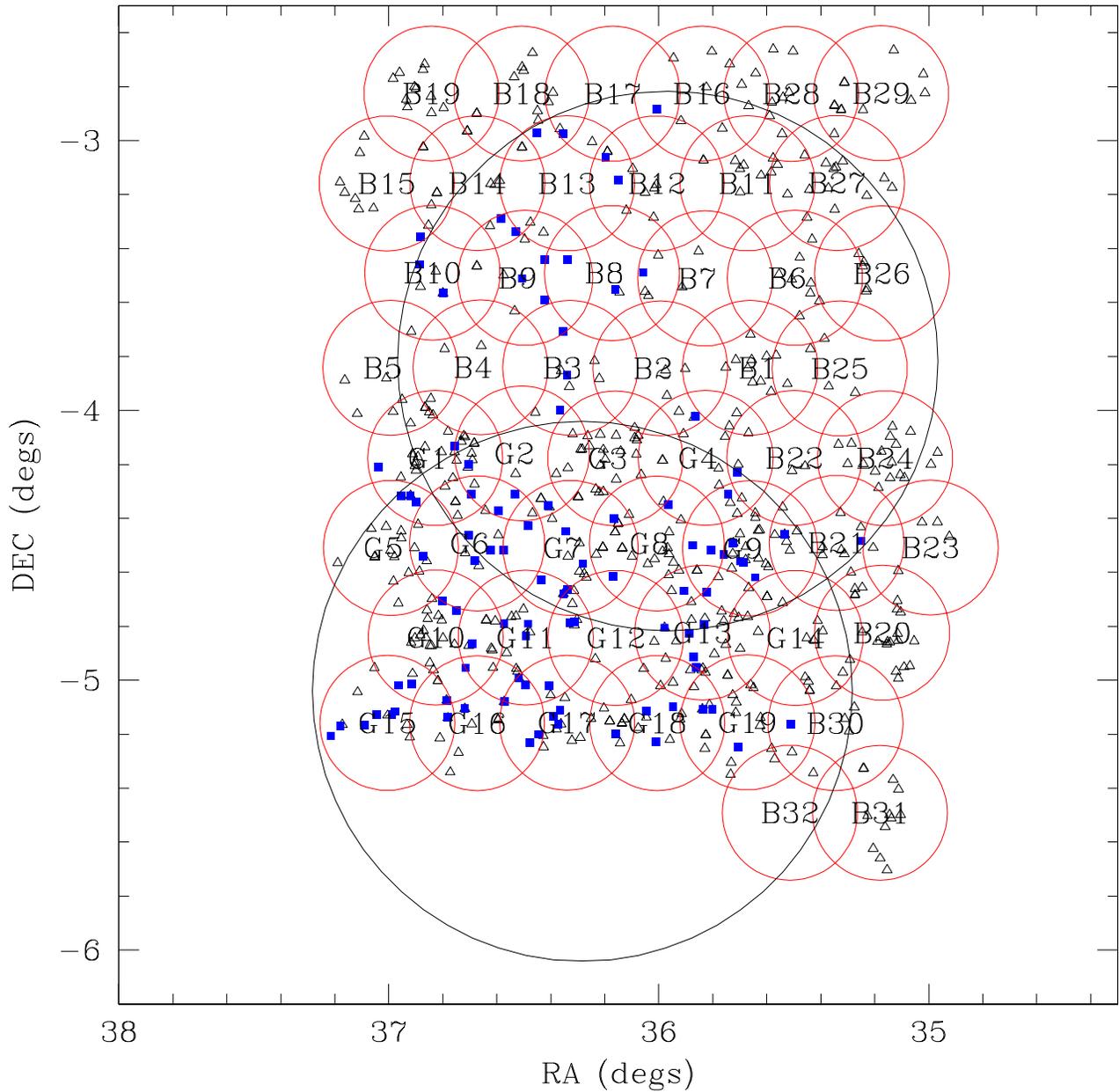}
  \caption{The layout of the 51 X-ray pointings of the XMM-LSS survey is such that
  most adjacent pointings overlap beyond 10 arcmin from their respective optical axis centers. The filled
  squares correspond to the 99 spectroscopically identified X-ray sources for this analysis, while the empty triangles
  correspond to the remaining 513 X-ray sources.
  Finally, the two large overlapping circles correspond to the two 2dF fields used in the spectral identification process.}
  \label{xmmlss}
\end{figure*}

\begin{figure*}
\centering
  \includegraphics[angle=0,width=9cm]{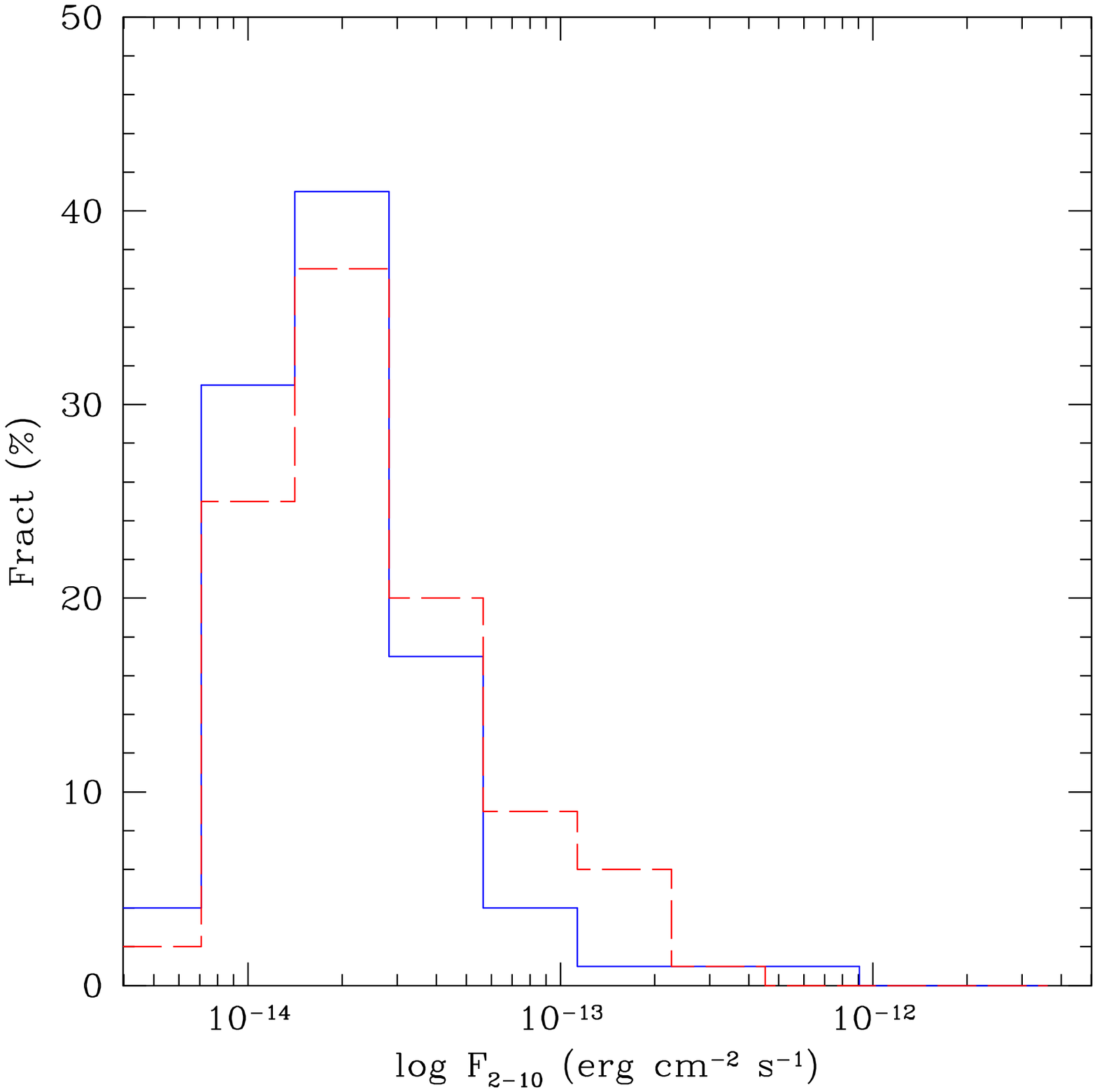}
  \includegraphics[angle=0,width=9cm]{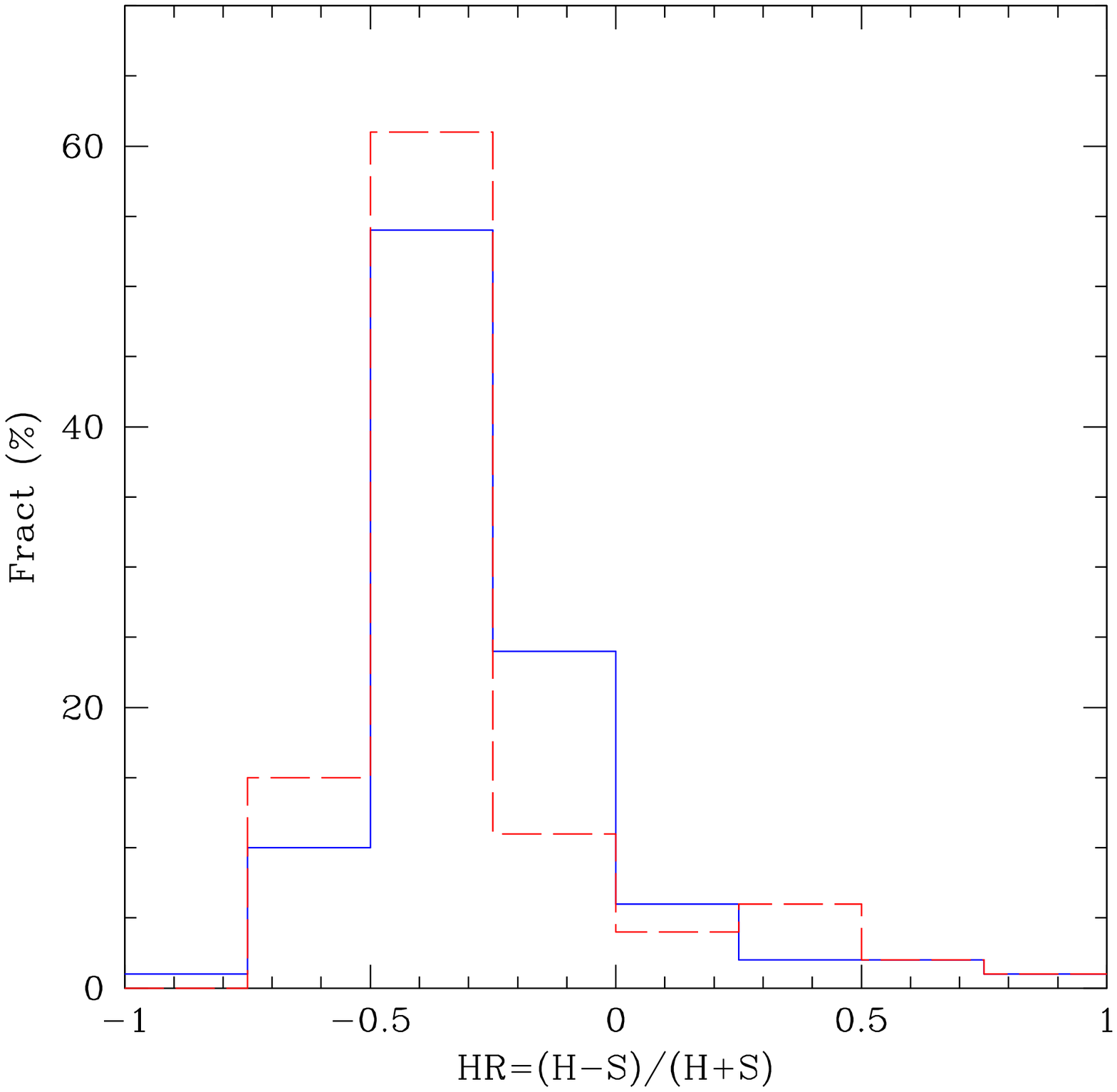}
  \caption{Left panel : normalized 2-10 keV band flux distribution for the 513 still optically unidentified X-ray sources (solid line)
  compared to the distribution of the sample of 99 spectroscopically selected X-ray point like sources (long dashed line).
  Right panel : Normalized hardness ratio distribution for the 513 still optically unidentified X-ray sources (solid line)
  compared to that of the sample of 99 optically selected X-ray point like sources (long dashed line). }
  \label{histofluxall}
\end{figure*}

\begin{figure*}
\centering
  \includegraphics[angle=0,width=9cm]{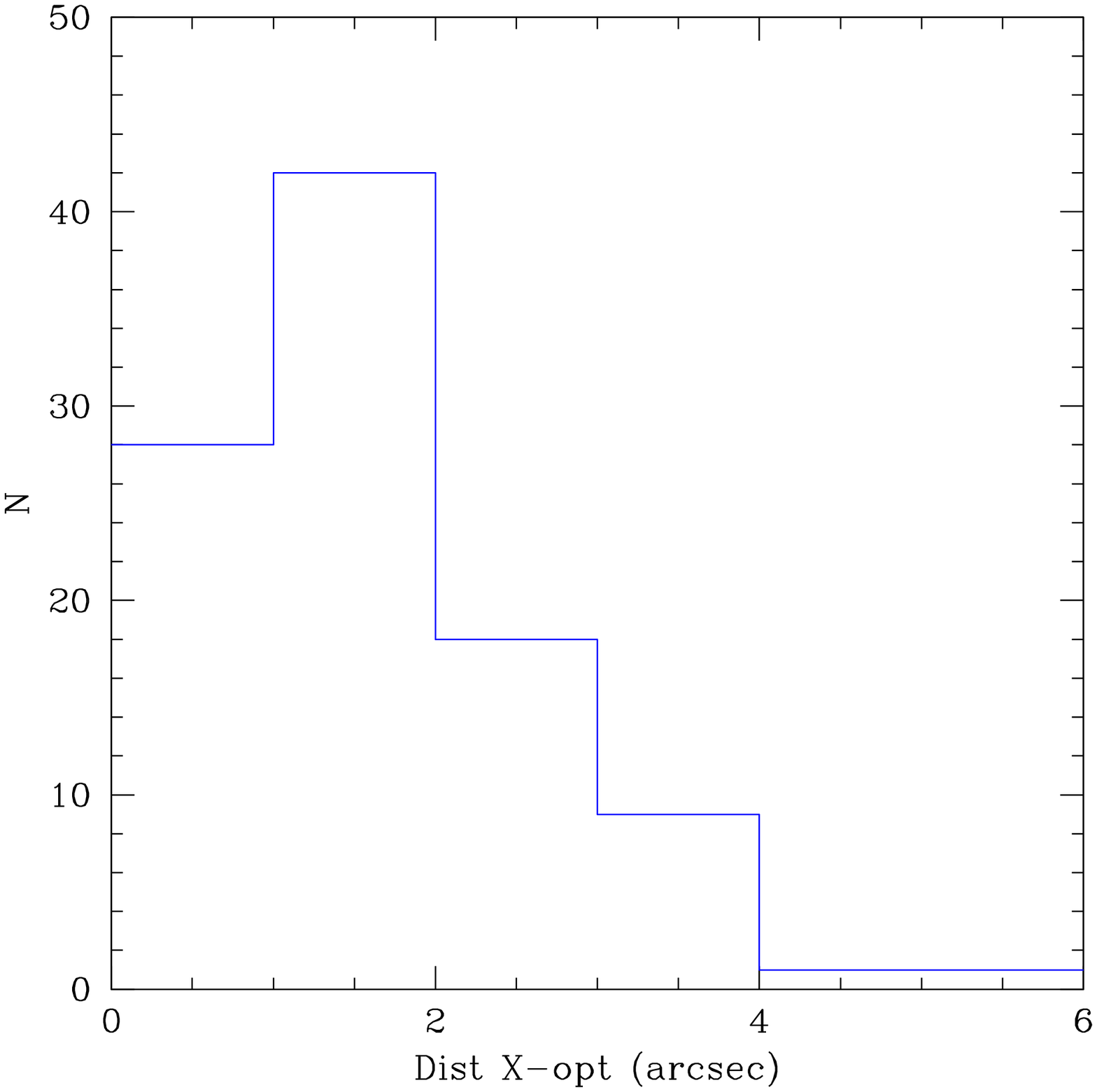}
  \includegraphics[angle=0,width=9cm]{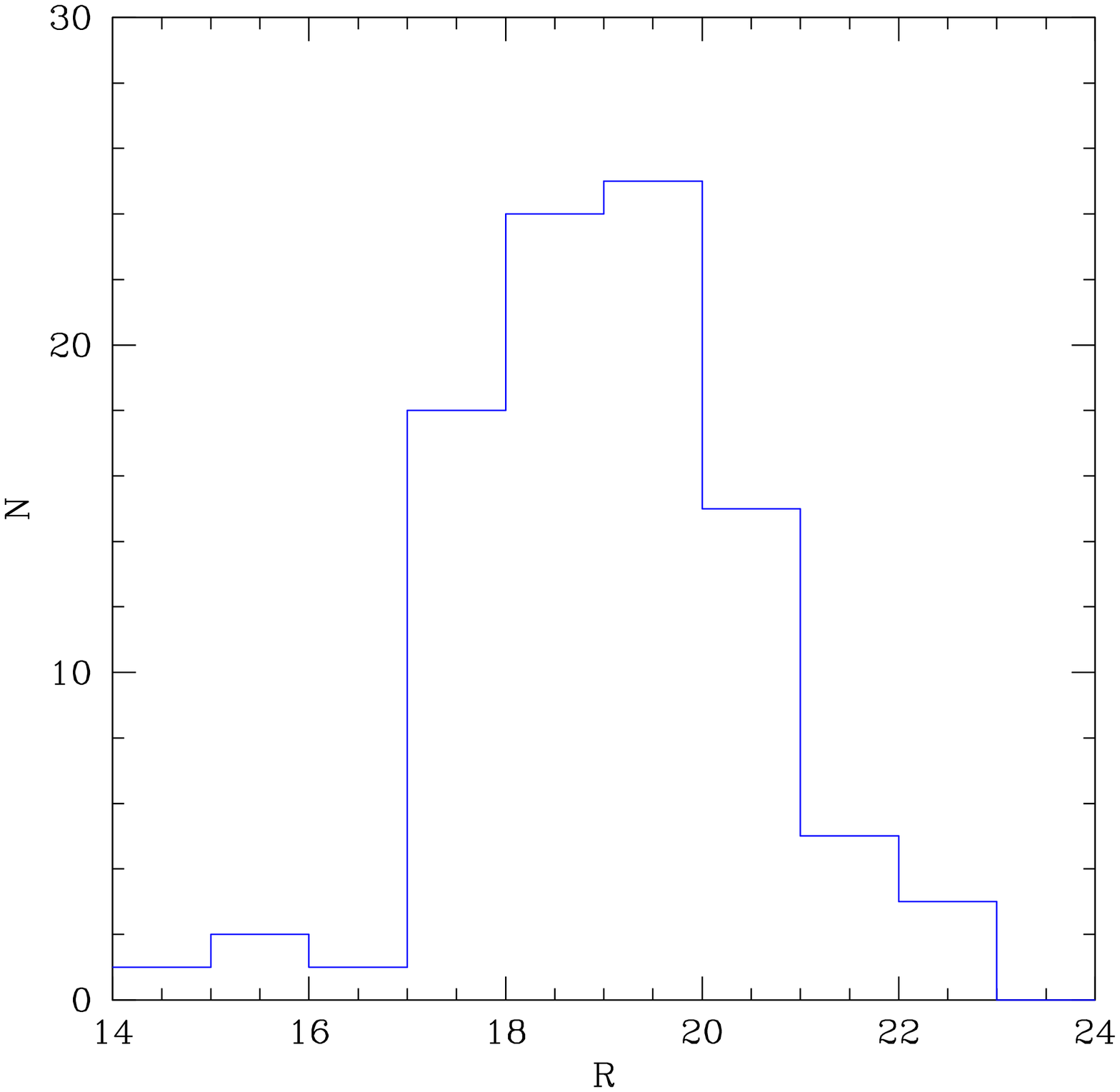}
  \caption{Left panel : histogram of the distribution of the separation between
  each X-ray source and its optical counterpart, for the 99 spectroscopically identified
  X-ray sources. Right panel : $R$ band magnitude distribution of the 94 optical counterparts of our
  sample of X-ray sources for which an $R$ band magnitude is available. For most of
  the objects, the magnitude is red magnitude from UKST plate scans, while the others
  are $R$ magnitudes from the VVDS. }  \label{histodist}
\end{figure*}

\end{document}